\documentclass[11pt]{amsart}

\usepackage{amscd}
\usepackage{amsmath}
\usepackage{graphicx}
\usepackage{amsfonts}
\usepackage{amssymb}
\usepackage{makecell}
\usepackage{algorithm}
\usepackage{algorithmic}
\usepackage{epstopdf}
\usepackage{epsfig}
\usepackage{xcolor}
\usepackage{booktabs}
\usepackage{multirow}
\usepackage{hyperref}

\usepackage{float}
\usepackage{array}
\usepackage{booktabs}
\usepackage{adjustbox} 
\usepackage{multirow}
\usepackage{algorithm}
\usepackage{algorithmic}
\usepackage{bm}

\newcolumntype{C}[1]{>{\centering\arraybackslash}m{#1}}

\begin{document}

\title[Multipartite correlation measures]
{Multipartite correlation measures and framework for multipartite quantum resources theory
}

\author{Taotao Yan}\email{18834809565@163.com}
\author{Jinchuan Hou} \email{jinchuanhou@aliyun.com}

\address{College of Mathematics, Taiyuan University of Technology, Taiyuan, 030024, P. R.
China}
\author{Xiaofei Qi}\email{xiaofeiqisxu@aliyun.com}
\address{School of Mathematical Science, Shanxi University, Taiyuan 030006,
P. R. China; Key Laboratory of Complex Systems and Data Science of Ministry of Education,
Shanxi University, Taiyuan  030006,  Shanxi, China}
\author{ Kan He}\email{hekan@tyut.edu.cn}
\address{College of Mathematics, Taiyuan University of Technology, Taiyuan, 030024, P. R.
China}

\begin{abstract}
In recent years, it has been recognized that properties of multipartite physical systems, such as genuine multipartite entanglement, can be considered as important resources for quantum information and other areas of physics. However, the current framework of multipartite quantum resource theory is flawed. In this paper, we propose a more reasonable  framework for multipartite quantum resource theory with a particular focus on axiomatic definition for true measures of multipartite quantum correlations (MQC)  that regulates how to measure the correlation in part systems (the unification condition) and how to describe the requirement from many-body resource theory that the correlation hold by part system does not exceed that of the entire system (the hierarchy condition).  We find that, due to the inherent characteristics of MQCs, the true measures of different MQCs exhibit distinct hierarchy conditions.  Based on this framework, we verify that multipartite entanglement, $k$-entanglement, $k$-partite entanglement, multipartite non-PPT, multipartite coherence, multipartite imaginarity,  multipartite multi-mode Gaussian non-product correlation, multipartite multi-mode Gaussian imaginarity, and multipartite single-mode Gaussian coherence are all symmetric multipartite quantum resources. We also show that, multipartite steering is an asymmetric multipartite quantum resource. Finally,  the monogamy relations for true measures of symmetric MQCs are  discussed.

\end{abstract}

\maketitle

\section{Introduction}

Multipartite quantum correlations (MQCs) are an essential component of modern physics and serve as a key driving force behind quantum communication and computational technologies. Specifically, these MQCs  play a crucial role in the fields of quantum information processing and quantum computing \cite{MI}, as well as in quantum phase transitions and the detection of other cooperative quantum phenomena in various interacting quantum multipartite systems \cite{LRA,VJW}.  Thus, every MQC is potentially a kind of quantum resource.

The current quantum resource theory requires that a quantum resource contains three basic elements: free states, free operations, and a resource measure that is non-increasing  under free operations \cite{FG,EG,GG}. Non-free states then are the resource states.  By the known results so far, there are three kinds of quantum resources. The first one is the class of quantum resources of which the free states are defined in  single particle systems, such as the quantum coherence \cite{AGB} and the quantum imaginarity \cite{DVS}. The second kind consists of  those for which the resource  states are bipartite quantum  states for some correlation such as the entanglement \cite{RPM}, the nonlocality \cite{DVJI} and the steering \cite{RLA}. The third kind consists of those for which the resource states are certain multipartite quantum correlated states.

In \cite{FG}, Brandao et al. provided a general framework for multipartite resource theory. In this framework, the set of free states should satisfy some physical motivations. Specifically, the set of free states should be closed under tensor products, as well as under partial traces and permutations of spatially separated subsystems. Additionally, the set of free states should be a closed convex set. Free operations cannot generate resources, that is, they cannot convert free states into resource states. The MQC measures are nonnegative functionals that satisfy   faithfulness and  non-increasing trend under free operations.

 Recently, for the free states, researchers discovered the existence of quantum resources that extend beyond convexity and are not closed under permutations of spatially separated subsystems in \cite{RJR}. For example, quantum steering and discord exhibit these properties. Therefore, the assumptions that the set of free states must be convex and closed under permutations of spatially separated subsystems are not necessary.

As to the MQC measures, the situation becomes more complex. As resource measures, they should obey more requirements especially from many-body resource theory. The  properties such as faithfulness and the monotonicity under free operations are sufficient for bipartite QC measures but not enough for MQC measures of $n$-partite systems with $n\geq 3$ because at least two issues are not solved: How to measure the resource in part systems and how to describe the regulation that the resource hold by part system can never exceed that of entire system?

\if false
 MQC measures  become more complex and require additional basic conditions such as the unification condition and the hierarchy condition imposed by resource allocation theory. Therefore, in this paper, we propose an axiomatic definition of   MQC measures, thereby revisiting the theory of multipartite quantum resources. To distinguish the current MQC measures that only satisfy faithfulness and non-increasing trend under free operations, we call the MQC  measures that satisfy the axioms  {\it true}   MQC measures.\fi

In the past two decades, continuous-variable (CV) systems have garnered widespread attention from both theoretical and experimental perspectives. Among them, Gaussian state is uniquely characterized by its mean vector and covariance matrix. Moreover, the challenges associated with infinite-dimensional aspects of Gaussian states can often be reduced to problems involving finite-dimensional matrix and vector algebra, thereby significantly simplifying the analysis.  Currently, researches on Gaussian quantum resource theory were active, such as the bipartite Gaussian steering \cite{TJJ}, and Gaussian imaginarity \cite{XJW}.

For multipartite multi-mode Gaussian correlations, a framework of Gaussian resource theory was proposed in \cite{LBX}.

In this framework, the set of free states must satisfy several conditions: it should be invariant under displacement operations, closed under tensor products, closed under partial traces and permutations of spatially separated subsystems, and it should form a closed set. Additionally, the set of covariance matrices corresponding to Gaussian free states is upward closed. Recently, it has been discovered that certain multipartite Gaussian resources do not satisfy the condition of being invariant under displacement operations, and the set of covariance matrices corresponding to the set of free states is not upward closed, for example, considering multipartite single-mode Gaussian coherence \cite{HHQ24}. Additionally, the set of asymmetric multipartite Gaussian free states is not closed under permutations of spatially separated subsystems, as demonstrated in the case of multipartite Gaussian steering \cite{YIG}.

Free operations are Gaussian channels that convert any Gaussian free state into another Gaussian free state.
And Gaussian quantum correlation measures should satisfy  faithfulness and non-increasing trend under free operations.

However, motivated by multipartite multi-mode Gaussian non-product correlation \cite{HLQ22}, a true multipartite Gaussian quantum correlation  measure should meet additional  conditions as required by resource allocation theory. Consequently, revisiting the theory of multipartite Gaussian quantum resources is also necessary.

In summary, there are two issues in the existing quantum resource theory when considering MQCs. One is the requirement of free states, and the other is the definition of correlation measure for MQCs.

In this paper, we establish a more reasonable framework for multipartite quantum resource theory, primarily by proposing an axiomatic definition of {\it true} MQC measures that solves the issues existing in the current framework. The axiomatic definition of  true MQC measures should satisfy at lest four principles, namely,
 faithfulness, monotonicity  under free operations, unification condition for measuring MQC in part systems  and hierarchy condition for ensuring MQC in part system does not exceed that in whole system. For several MQCs that have been extensively studied, we provide exactly true MQC measures that meet our axiomatic definition.
  Additionally, we provide a precise definition of the monogamy relations for true measures of symmetric MQCs.

This paper is organized as follows. In Section 2, we establish a framework for multipartite quantum resource theory, focusing on defining the principles that the true MQC measures should satisfy.  In Section 3, we discuss the symmetric MQCs, and show  that the multipartite entanglement, $k$-entanglement,  $k$-partite entanglement, multipartite non-PPT, multipartite coherence, multipartite imaginarity, multipartite multi-mode Gaussian non-product correlation, multipartite multi-mode Gaussian imaginarity and multipartite single-mode Gaussian coherence are all multipartite quantum resources. In Section 4, we discuss asymmetric MQCs, and
prove that the multipartite steering is an asymmetric multipartite quantum resource. In Section 5, we discuss the monogamy relations for true measures of symmetric MQCs. Section 6 is a short conclusion.

\section{A framework of multipartite quantum resource theory}

In this section, we revisit the three basic elements required for multipartite quantum resource theory: free states, free operations, and true  measures of multipartite resource, aiming to establish a more precise framework of MQC resource theory suitable to many-body systems.

For any $n$-partite system $H=H_1\otimes H_2\otimes \cdots\otimes H_n$ of any dimension, let ${\mathcal S}(H)$ denote the set of all $n$-partite states in  system $H$.

\subsection{Free states}
 The set of free states forms a closed subset of ${\mathcal S}(H)$.  It is also important to note that, for multipartite Gaussian quantum resource theory  in CV system, the collection of Gaussian free states should also form a closed set.

\subsection{Free operations}

Generally, we  take a suitable subset of all  operations   that always sends free states into free states as the free operations under consideration. Similarly, for multipartite Gaussian quantum resource theory in CV system, the free operations form a subset of all   Gaussian channels that convert any Gaussian free state into   Gaussian free state.

\subsection{The true MQC measures}
For a bipartite quantum correlation (BiQC), a measure of this BiQC is a non-negative function $\mathcal C^{(2)}$ on states  such that (1) (Faithfulness)  $\mathcal C^{(2)}(\rho)>0$ if and only if $\rho$ is a BiQC state; (2) (Monotonicity under free operations) For any free channel $\Phi$ and any state $\rho$, we always have $\mathcal C^{(2)}(\Phi(\rho))\leq \mathcal C^{(2)}(\rho)$ \cite{FG}. This aligns with the conditions that a multipartite  measure  $\mathcal C^{(n)}$ is required to satisfy within the current framework of multipartite resource theory.

However, for multipartite systems, the situation becomes increasingly intricate. For a MQC measure  $\mathcal C^{(n)}$, the  faithfulness and non-increasing trend  under free operations are not enough. For example, by the principle of resource allocation, the measure of the part systems cannot exceed the total measure available to the entire system. This property was described as {\it the hierarchy condition} for multipartite entanglement measures in multipartite systems \cite{GY20} and multipartite multi-mode Gaussian non-product measure in multipartite multi-mode CV systems \cite{HLQ22}. This property further requires that, one has to measure the resource in any part system. Thus, if $\mathcal C^{(n)}$ is a  MQC measure, it is not enough to consider $\mathcal C^{(n)}$ itself only, instead, one must consider a sequence $\{\mathcal C^{(r)}\}_{r=2}^n$
as a family and each $\mathcal C^{(r)} $ is defined in the same way of $\mathcal C^{(n)}$ and gets well along with each other, that is, causing no confusions when used them to measure the MQC in part system. This requirement was described as {\it the unification condition} for MQC measures \cite{HLQ22}.

Recall that,  $P=P_1|P_2|\cdots|P_m$ is called a sub-repartition of $\{1,2,\cdots,n\}$ if $P_j\subset\{1,2,\cdots,n\}$, $P_j\cap P_i=\emptyset$ whenever $i\not= j$ and $\bigcup_{j=1}^m P_j\subseteq \{1,2,\cdots,n\}$. $P$ determines a part system $H_P=H_{P_1}\otimes H_{P_2}\otimes \cdots\otimes H_{P_m}$ of the whole system $H=H_1\otimes H_2\otimes\cdots\otimes H_n$. Denote by $\mathcal{SP}_n(H)$ the set of all sub-repartitions of $\{1,2,\ldots,n\}$.
More generally, the hierarchy condition is in fact a natural requirement from the theory of resource allocation as well as the feature of MQC  itself, which states that, for any sub-repartitions $Q=Q_1|Q_2|\cdots|Q_r, P=P_1|P_2|\cdots|P_m\in \mathcal{SP}_n$ of $\{1,2,\cdots,n\}$, if the hierarchy of $Q$ is lower than  the hierarchy of $P$, denoted by $Q\preccurlyeq_{\rm MQC} P$, then, for any state $\rho\in\mathcal S(H_1\otimes H_2\otimes\cdots\otimes H_n)$, regarding its reduced states $\rho_Q\in{\mathcal S}(H_{Q_1}\otimes H_{Q_2}\otimes\cdots\otimes H_{Q_r})$ as a $r$-partite state and $\rho_P\in{\mathcal S}(H_{P_1}\otimes H_{P_2}\otimes\cdots\otimes H_{P_m})$ as an $m$-partite state, one always has  $\mathcal C^{(r)}(\rho_Q, Q)\leq \mathcal C^{(m)}(\rho_P,P)$, that is, the correlation MQC contained in part system $H_Q$  can not exceed the correlation MQC shared by $H_P$ whenever   $Q\preccurlyeq_{\rm MQC} P$.

Note that the meaning of $Q\preccurlyeq_{\rm MQC} P$ for any two sub-repartitions $P,Q\in\mathcal{SP}_n$ may vary according to different MQCs, which will be discussed in detail in the following sections. If the hierarchy of $Q$ is lower than the hierarchy of $P$ with respect to MQC, that is, if $Q\preccurlyeq_{\rm MQC} P$, we also say that $Q$ is coarser than $P$ or, equivalently, that $P$ is finer than $Q$ with respect to MQC. Generally, $\preccurlyeq_{\rm MQC} $ is a partial order in $\mathcal{SP}_n$.

By the discussion above,  we propose an axiomatic definition of MQC measures.

{\bf Definition 2.1.}  A  quantification $\mathcal C^{(n)}$ of a $n$-partite quantum correlation $\mathcal C$  for system $H=H_1\otimes H_2\otimes \cdots\otimes H_n$ is a true MQC measure if it satisfies the following conditions:

{\bf (MQCM1)} (Non-negativity and faithfulness) $\mathcal C^{(n)}(\rho)\geq 0$ for all $\rho\in{\mathcal S}(H_1\otimes H_2\otimes \cdots\otimes H_n)$ and $\mathcal C^{(n)}(\rho)= 0$ if and only if $\rho$  contains no correlation $\mathcal C$.

{\bf (MQCM2)} (Monotonicity under free operations) For any free operation $\Phi$,  ${\mathcal C}^{(n)}(\Phi(\rho))\leq {\mathcal C}^{(n)}(\rho)$ holds for all $\rho\in {\mathcal S}(H_1\otimes H_2\otimes \cdots\otimes H_n)$.

{\bf (MQCM3)} (Unification condition) $\{{\mathcal C}^{(r)}\}_{r=2}^n$ are defined in the same way and can be used to measure MQC $\mathcal C$ in any part systems without causing confusion.

{\bf (MQCM4)} (Hierarchy condition) There is a hierarchy order $\preccurlyeq_{\mathcal C}$  in $\mathcal{SP}_n$ determined by $\mathcal C$ such that, for $P=P_1|P_2|\cdots|P_m, Q=Q_1|Q_2|\cdots|Q_r\in\mathcal{SP}_n$, if $Q\preccurlyeq_{\mathcal C} P$ (that is, the hierarchy of $Q$ is lower than the hierarchy of $P$ with respect to $\mathcal C$), then ${\mathcal C}^{(r)}(\rho_Q, Q)\leq {\mathcal C}^{(m)}(\rho_P,P)$ holds for all $\rho\in {\mathcal S}(H_1\otimes H_2\otimes \cdots\otimes H_n)$, where $\rho_P$ and $ \rho_Q$ are respectively the reduced states of $\rho$ to the subsystems $H_P$ and $H_Q$, regarded as $r$-partite state and $m$-partite state.

We call $\mathcal C^{(n)}$ satisfying the axioms (MQCM1)-(MQCM4)  a {\it true} multipartite measure of $\mathcal C$, to distinguish it from the current  measures satisfying only (MQCM1)-(MQCM2).
The conditions (MQCM1)-(MQCM4) are the most basic requirements for a true multipartite measure. For instance, if the MQC $\mathcal C$ is symmetric with respect to the subsystems, one should require that a true multipartite measure is invariant under order changing of  subsystems.

Therefore, it is reasonable to propose a general framework for multipartite quantum resources.

{\bf Definition 2.2.} A MQC $\mathcal C$ is a multipartite quantum resource if it satisfies the following conditions:

(1) The set of non-MQC states forms a closed set, and the non-MQC states are treated as free states. The MQC states are thus the resource states.

(2) A suitable subset of all channels that map non-MQC states to non-MQC states (referred to as a non-MQC channel) is chosen and considered as free operations.

(3) There exists a true MQC measure of $\mathcal C$ satisfying the conditions (MQCM1)-(MQCM4).

The notion of multipartite Gaussian quantum resource can be defined similarly for multipartite multi-mode  CV systems.

So, to check whether or not a MQC is a multipartite quantum resource, the key is the existence of a true MQC measure.

In the next two sections we show that several  widely studied and applied MQCs are multipartite quantum resources in the sense of this new framework by presenting  true MQC measures.

\section{Symmetric MQCs as multipartite quantum resources }

In this section we present several symmetric MQCs that are multipartite quantum resources. Recall that a MQC $\mathcal C$ is symmetric if the property  that a state has $\mathcal C$ is invariant under changing the order of subsystems.

\subsection{Completely symmetric multipartite quantum resources}
A MQC is said to be completely symmetric if it can be defined for any $n$-partite system $H=H_1\otimes H_2\otimes \cdots\otimes H_n$ of any dimension and it is symmetric with respect to the subsystems of any partitions. This means that, for any partition $P=P_1|P_2|\cdots|P_m$  of $\{1,2,\cdots,n\}$, $2\leq m\leq n$, and any $\rho\in {\mathcal S}(H_1\otimes H_2\otimes \cdots\otimes H_n)$,    $\rho$ is MQC as an $m$-partite state in ${\mathcal S}(H_{P_1}\otimes H_{P_2}\otimes \cdots\otimes H_{P_m})$ implies $\rho^{\pi,P}\in{\mathcal S}(H_{P_{\pi(1)}}\otimes H_{P_{\pi(2)}}\otimes \cdots\otimes H_{P_{\pi(m)}})$ is MQC  for any permutation $\pi$ of $(1,2,\cdots,m)$, where $\rho^{\pi,P}$ is obtained from $\rho$ by changing the orders of subsystems $P_j$ according to $\pi$.

Clearly, multipartite entanglement ($\rho$ is entangled if and only if $\rho$ is not fully separable) is a completely  symmetric MQC.

In this situation, except for (MQCM1)-(MQCM4), a true MQC measure $\mathcal C^{(n)}$ should satisfy one more condition:

 {\bf (MQCM5)} (Symmetry) $\mathcal C^{(n)}$ is invariant under permutation of subsystems, that is, $\mathcal C^{(n)}(\rho^\pi) =\mathcal C^{(n)}(\rho)$ for any permutation $\pi$ of $(1,2,\cdots,n)$.

Usually, the conditions (MQCM1)-(MQCM3) and (MQCM5) are well understood but (MQCM4) is not.  What does exactly the condition (MQCM4) mean for a completely symmetric MQC?
This is equivalent to asking what it means for the hierarchy of a sub-repartition $Q=Q_1|Q_2|\cdots|Q_r$ to be lower than that of a sub-repartition $P=P_1|P_2|\cdots|P_m$ with respect to this MQC.

Assume $2\leq r<n$ and consider the situation $Q=1|2|\cdots|r$ and $P=1|2|\cdots|n$. Thus, the  subsystems are $H_Q=H_1\otimes H_2\otimes \cdots\otimes H_r$ and $H_P=H=H_1\otimes H_2\otimes\cdots\otimes H_n$.  Since the theory of resource allocation requires that  the correlation held by part of the system cannot exceed the whole correlation of the system, we should have, for any $\rho\in {\mathcal S}(H)$, $\mathcal C^{(r)}(\rho_Q)\leq \mathcal C^{(n)}(\rho)$. This suggests the following basic relations between sub-repartitions.

{\bf (a)}  For any sub-repartitions $P=P_1|P_2|\cdots|P_m$ and $Q=Q_1|Q_2|\cdots |Q_r$ of $\{1,2,\cdots, n\}$, if $\{Q_i\}_{i=1}^r\subseteq\{P_j\}_{j=1}^m$, that is,  for each $i=1,2,\cdots, r$, $Q_i=P_{j_i}$ for some $ j_i$, we say that  $Q$ is coarser than $P$ in type (a), denoted by $Q\preccurlyeq^a P$.

But $Q\preccurlyeq^a P$ cannot cover all situations. For example, let $Q=Q_1|Q_2|\cdots |Q_r$ be a $r$-partition of $\{1,2,\cdots,n\}$ and $P=1|2|\cdots|n$. In this scenario, if the MQC $\mathcal C$ is multipartite entanglement, for any $\rho\in{\mathcal S}(H)$, regarding $\rho\in {\mathcal S}(H_{Q_1}\otimes H_{Q_2}\otimes\cdots\otimes H_{Q_r})$, the value of ${\mathcal C}^{(r)}(\rho,Q)$ can be regarded as the entanglement shared by subgroups $\{Q_1,Q_2,\cdots,Q_r\}$, which should not exceed  the entanglement shared by entrie system $H=H_1\otimes H_2\otimes\cdots\otimes H_n$. Therefore, we should deduce that the hierarchy of $Q$ is lower than the hierarchy of $P=1|2|\cdots|n$ and ${\mathcal C}^{(r)}(\rho,Q)\leq {\mathcal C}^{(n)}(\rho,P)$. Furthermore, if we kick out some members from some $Q_i$ of $Q=Q_1|Q_2|\cdots|Q_r$ and get another sub-repartition $R=R_1|R_2|\cdots|R_r$, which means $\emptyset\not=R_i\subseteq Q_i$ for each $i=1,2,\cdots,r$, it is natural to expect that the entanglement imposed by $R$ cannot exceed the entanglement imposed by $Q$, that is, ${\mathcal C}^{(r)}(\rho_R,R)\leq {\mathcal C}^{(r)}(\rho_Q,Q)$ for any $\rho\in{\mathcal S}(H_1\otimes H_2\otimes\cdots\otimes H_n)$. These suggest the following

{\bf (b)} For any sub-repartitions $P=P_1|P_2|\cdots|P_m$ and $Q=Q_1|Q_2|\cdots |Q_r$ of $\{1,2,\cdots, n\}$, if  $Q$ is a partition of $P$, that is, for each $i=1,2,\cdots, r$, there exist $j_{1,i}, \cdots, j_{s_i,i}$ such that $Q_i=\bigcup_{t=1}^{s_i} P_{j_{t,i}}$, we say that  $Q$ is coarser than  $P$ in type (b), denoted by $Q\preccurlyeq^b P$.

{\bf (c)} For any sub-repartitions $P=P_1|P_2|\cdots|P_m$ and $Q=Q_1|Q_2|\cdots |Q_r$ of $\{1,2,\cdots, n\}$, if $r=m$ and  $Q_i$ is obtained by removing out some elements from $P_{j_i}$, that is, $\emptyset\not= Q_i\subseteq P_{j_i}$ and ${j_i}\not=j_k$ whenever $k\not= i$, $i=1,2,\cdots, m$, we say that  $Q$ is coarser than  $P$ in type (c),  denoted by $Q\preccurlyeq^c P$.

Generally, we say that
sub-repartition $Q$ is coarser than  sub-repartition  $P$, denoted by $Q\preccurlyeq P$, if there exist finitely many  $R_1,R_2,\cdots, R_t\in\mathcal{SP}_n$ such that
\begin{equation}\label{eq3.1}  Q\preccurlyeq^{x_1} R_1\preccurlyeq^{x_2} R_2 \preccurlyeq^{x_3}\cdots \preccurlyeq^{x_t}R_t\preccurlyeq^{x_{t+1}} P,\tag{3.1}\end{equation}
where $x_1,x_2,\cdots,x_t,x_{t+1}\in\{a,b,c\}$.
\if false
Let
 $$\mathcal{SP}_{n}=\{P: P \ \mbox{\rm is a sub-repartition of}\ \{1,2,\cdots,n\}\}.$$\fi Clearly, ``$\preccurlyeq^x$'' is a partial order relation of the set $\mathcal{SP}_{n}$ for any $x\in\{a,b,c\}$, and consequently, ``$\preccurlyeq$'' is also a partial order relation of the set $\mathcal{SP}_{n}$.

For a completely symmetric MQC $\mathcal C$, the hierarchy relation $\preccurlyeq_{\mathcal C}$ between sub-repartitions in the rule (MQCM4) can be defined with the help of the above three basic relations $\preccurlyeq^x$, $x\in\{a,b,c\}$.

\if false
Therefore, the condition (MQCM4) for completely symmetric MQC measure $\mathcal C^{(n)}$ can be rewritten as (CS-MQCM4):

(CS-MQCM4) (Hierarchy condition) $\mathcal C^{(n)}$ satisfies the following three conditions: for any $P=P_1|P_2|\cdots|P_m\in\mathcal{SP}_{n}$, $Q=Q_1|Q_2|\cdots|Q_r\in\mathcal{SP}_{n}$, and any $\rho\in {\mathcal S}(H_1\otimes H_2\otimes \cdots\otimes H_n)$,

(CS-MQCM4a) $Q\preccurlyeq^a P$ implies $\mathcal C^{(r)}(\rho_Q, Q)\leq \mathcal C^{(m)}(\rho_P, P)$;

(CS-MQCM4b) $Q\preccurlyeq^b P$ implies $\mathcal C^{(r)}(\rho_Q, Q)\leq \mathcal C^{(m)}(\rho_P, P)$;

(CS-MQCM4c) $Q\preccurlyeq^c P$ implies $\mathcal C^{(m)}(\rho_Q, Q)\leq \mathcal C^{(m)}(\rho_P, P)$.
\fi

\subsection{Entanglements in multipartite systems}

We demonstrate that the multipartite entanglement, $k$-entanglement and $k$-partite entanglement are  completely symmetric multipartite quantum resources.

 {\bf Example 3.1.} Multipartite entanglement  is a completely symmetric multipartite quantum resource.


 For multipartite entanglement, the free states are (fully) separable states, and the free operations are all local operations and classical communications (LOCCs). So, as a multipartite quantum resource, it should have a true multipartite entanglement measure satisfying the conditions (MQCM1)-(MQCM5).

For $n=3$, several tripartite entanglement measures such as $E_{f}^{(3)}, C^{(3)}$ and $T_{q}^{(3)}$ were proposed in \cite{GY20}, where
 $$E_{f}^{(3)}(|\psi\rangle)=\frac{1}{2}[S(\rho_{1})+S(\rho_{2})+S(\rho_{3})],$$
 $$C^{(3)}(|\psi\rangle)=[3-{\rm Tr}(\rho_{1})^{2}-{\rm Tr}(\rho_{2})^{2}-{\rm Tr}(\rho_{3})^{2}]^{\frac{1}{2}},$$
 $$T_{q}^{(3)}(|\psi\rangle)=\frac{1}{2}[T_{q}(\rho_{1})+T_{q}(\rho_{2})+T_{q}(\rho_{3})], q>1$$
 for pure state $|\psi\rangle\in H_{1}\otimes H_{2}\otimes H_{3}$, and  for any $i=1,2,3$, $i^{c}=\{1,2,3\}\backslash \{i\}$, $\rho_{i}={\rm Tr}_{i^{c}}(|\psi\rangle\langle\psi|)$. Recall that, $S(\rho)=-{\rm Tr}(\rho{\rm log}\rho)$ is the von Neumann entropy and  $T_{q}(\rho)=(1-q)^{-1}[{\rm Tr}(\rho^{q})-1]$ is the Tsallis $q$-entropy. Then the value for mixed state is defined via  the convex-roof extension, that is,
\begin{equation}\mathcal{E}^{(3)}(\rho)=\mathop{\min}\limits_{\{p_{i},|\psi_{i}\rangle\}}\mathop{\sum}
 \limits_{i}p_{i}\mathcal{E}^{(3)}(|\psi_{i}\rangle) \tag{3.2}\end{equation}
 for any mixed state $\rho\in {\mathcal S}(H_{1}\otimes H_{2}\otimes H_{3})$. Here the minimum is taken over all possible pure decompositions $\rho=\mathop{\sum}\limits_{i}p_{i}|\psi_{i}\rangle\langle\psi_{i}|$ and $\mathcal{E}^{(3)}$ is one of $E_{f}^{(3)}, C^{(3)}$ and $T_{q}^{(3)}$.

  In addition, for any
$\rho\in {\mathcal S}(H_{1}\otimes H_{2}\otimes H_{3})$,  $E_{f}^{(3)}, C^{(3)}$ and $T_{q}^{(3)}$ were shown to satisfy conditions (MQCM1)-(MQCM5) \cite{GY20}, where the hierarchy of a sub-repartition $Q\in\mathcal{SP}_{3}$ is lower than that of a sub-repartition $P\in\mathcal{SP}_{3}$ if and only if $Q\preccurlyeq P$. Thereby the above three measures are  true tripartite entanglement measures. It is evident that $E_{f}^{(3)}, C^{(3)}$ and $T_{q}^{(3)}$ can be generalized to arbitrary $n$-partite true entanglement measures $E_{f}^{(n)}, C^{(n)}$ and $T_{q}^{(n)}$, respectively. Therefore, the multipartite entanglement is a completely symmetric multipartite quantum resource.

An entanglement witness (EW) based true multipartite entanglement measure $E_w^{(n)}$  without convex roof extension  was proposed in \cite{GYHQH} for any $n$-partite systems, see Example 3.2 for case $k=n$.

 {\bf Example 3.2.}  For $2\leq k\leq n$, the $k$-entanglement in $n$-partite systems is a completely symmetric multipartite quantum resource.

 Now let us consider the $k$-entanglement (briefly, $k$-E) of $n$-partite systems.  Recall that, for $2\leq k\leq n$,  a pure state $|\psi\rangle\in H_1\otimes H_2\otimes\cdots\otimes H_n $ is said to be $k$-separable if there exists a $k$-partition $\{P_1,P_2,\cdots, P_k\}$ of $\{1,2,\cdots, n\}$ such that $|\psi\rangle$ is a product vector of the $k$-partite composite system $H_{P_1}\otimes H_{P_2}\otimes\cdots\otimes H_{P_k}$. A state $\rho\in {\mathcal S}(H_1\otimes H_2\otimes\cdots\otimes H_n )$ is said to be $k$-separable if it is a convex combination of $k$-separable pure states or the trace-norm limit of such convex combinations (this is the case for infinite-dimensional systems). Otherwise, it is said to be $k$-nonseparable or $k$-entangled. It is clear that $\rho$ is $n$-separable if and only if $\rho$ is fully separable.  The set ${\mathcal S}_k={\mathcal S}_k(H_1\otimes H_2\otimes\cdots\otimes H_n )$ of all $k$-separable states
 is a closed convex subset of ${\mathcal S}(H_1\otimes H_2\otimes\cdots\otimes H_n )$ and ${\mathcal S}_k\subset {\mathcal S}_l$ if $k>l$.  A state $\rho$ is   genuine entangled  if and only if  it is  $2$-entangled. Thus, the set of  entangled states is classified into $k$-entangled states, $k=n,n-1,\cdots, 3,2$. $k$-entanglement is also a completely symmetric MQC. Several $k$-entanglement measures via convex roof extension were proposed in \cite{GY24} which satisfy  the faithfulness ($k$-EM1), non-increasing trend under LOCC ($k$-EM2) and the symmetry ($k$-EM5), but none of these measures was proved to satisfy the hierarchy condition.

 Then, with $k$-separable states as free states and LOCCs as free operations, is the $k$-entanglement a multipartite quantum resource?

 Most recently, Ref.\cite{GYHQH} affirmatively addressed this question by proposing the first true $k$-entanglement measure that does not require a convex roof extension, satisfying conditions  ($k$-EM1)-($k$-EM5).

Not like the multipartite entanglement measures, the unification condition and the hierarchy condition for $k$-entanglement measures of $n$-partite systems are more complicated. Assume that $E^{(k,n)}$ is a true
$k$-entanglement measure, the unification condition requires that, for each $2\leq m\leq n$ and each $2\leq s\leq \min\{k,m\}$, the $s$-entanglement measure $E^{(s,m)}$  of $m$-partite system  should be defined.  So the unification for a $k$-entanglement measure $E^{(k,n)}$ reads as follows.

{\bf ($k$-EM3)} (Unification condition)
All $E^{(s,m)}$s are defined in the same way as $E^{(k,n)}$ so that for any sub-repartition $P=P_1|P_2|\ldots|P_m\in\mathcal{SP}_n$ and $2\leq s\leq \min\{k,m\}$, the $s$-entanglement $E^{(s,m)}(\rho_P,P)$ in $\rho_P\in\mathcal S(H_{P_1}\otimes H_{P_2}\otimes\cdots\otimes H_{P_m})$ can be measured for any state $\rho\in S(H_{1}\otimes H_{2}\otimes\cdots\otimes H_{n})$ without causing any confusions. \if false $\{E^{(s,m)}: 2\leq s\leq \min\{k,m\}, 2\leq m\leq n\}$ is gets along well  with each other.\fi

The hierarchy condition is a physical requirement from the theory of resource allocation. However, different from the situation of  entanglement (non-full separability) in Example 3.1, the hierarchy condition for $k$-entanglement measure is much different because $k$ is involved.  Roughly speaking, the hierarchy condition for $k$-entanglement measure ($2\leq k<n$) ensures that, if the hierarchy of a sub-repartition $Q=Q_1|Q_2|\cdots|Q_r$ with $2\leq k_1\leq r$ is lower than the hierarchy of a sub-repartition $P=P_1|P_2|\cdots|P_m$ with $2\leq k_2\leq m$ in terms of $k$-entanglement, denoted by $(k_1,Q)\preccurlyeq_{k-E} (k_2,P)$, then for any state $\rho\in \mathcal S(H_1\otimes H_2\otimes\cdots\otimes H_n)$,  the  $k_1$-entanglement of its reduced state $\rho_Q$ with respect to the part system $H_Q=H_{Q_1}\otimes H_{Q_2}\otimes\cdots\otimes H_{Q_r}$ can never exceed the  $k_2$-entanglement of its reduced state $\rho_P$ with respect to the part system $H_P=H_{P_1}\otimes H_{P_2}\otimes\cdots\otimes H_{P_m}$. Thus, to clarify the concept of the hierarchy condition for $k$-entanglement measure, one should first give the exact meaning of $(k_1,Q)\preccurlyeq_{k-E} (k_2,P)$. If $(k_1,Q)\preccurlyeq_{k-E} (k_2,P)$, we also say that $(k_1,Q)$ is coarser than $(k_2,P)$ with respect to $k$-entanglement, for simplicity.

By \cite{GYHQH}, for any $P=P_1|P_2|\cdots|P_m$, $Q=Q_1|Q_2|\cdots|Q_r\in\mathcal{SP}_{n}$,  there are three basic types in which $(k_{1}, Q)$ is coarser than  $(k_2, P)$ with respect to $k$-entanglement:

($k$-Ea) If $2\leq k_1\leq k_2\leq m$,  $ k_{1}\leq r$, $Q\preccurlyeq^a P$, and for any $k_2$ partition $R=R_1|R_2|\cdots|R_{k_2}$ of $P$,   $\#\{j: R_j\supseteq Q_{i_j}\ \mbox{\rm for some } Q_{i_j}\}\geq k_1$,  where $\#(F)$ stands for the cardinal number of the set $F$, we say that the hierarchy of $(k_1,Q)$ is lower than the hierarchy of $(k_2,P)$ in type (a), denoted by $(k_{1}, Q)\preccurlyeq^{a}_{k-E} (k_2, P)$;

($k$-Eb) If $2\leq k_1\leq k_2\leq m$,  $ k_{1}\leq r$, $Q\preccurlyeq^b P$, and for any $k_2$ partition $R=R_1|R_2|\cdots|R_{k_2}$ of $P$,  $\#\{j: R_j\supseteq Q_{i_j}\ \mbox{\rm for some } Q_{i_j}\}\geq k_1$, we say that the hierarchy of $(k_1,Q)$ is lower than the hierarchy of $(k_2,P)$ in type (b), denoted by $(k_{1}, Q)\preccurlyeq^{b}_{k-E} (k_2, P)$;

($k$-Ec) If $r=m$, $2\leq k_1\leq k_2\leq m$  and $Q\preccurlyeq^c P$, we say that the hierarchy of $(k_1,Q)$ is lower than the hierarchy of $(k_2,P)$ in type (c), denoted by $(k_{1}, Q)\preccurlyeq^{c}_{k-E} (k_2, P)$.

Then, for any $P=P_1|P_2|\cdots|P_m\in\mathcal{SP}_{n}$ and $Q=Q_1|Q_2|\cdots |Q_r\in\mathcal{SP}_{n}$, we say that the hierarchy of $(k_{1}, Q)$ is lower than the hierarchy of   $(k_2, P)$ with respect to $k$-entanglement, or, in other words, $(k_{1}, Q)$ is coarser than  $(k_2, P)$ with respect to $k$-entanglement, denoted by $(k_{1}, Q)\preccurlyeq_{k-E} (k_2, P)$,   if   there are some sub-repartitions $R_1,R_2,\cdots, R_t\in\mathcal{SP}_{n}$ and positive integers $r_1, r_2,\cdots, r_t$ such that
\begin{equation}(k_1,Q)\preccurlyeq^{x_1}_{k-E} (r_1,R_1)\preccurlyeq^{x_2}_{k-E} (r_2,R_2) \preccurlyeq^{x_3}_{k-E}\cdots \preccurlyeq^{x_t}_{k-E} (r_t,R_t)\preccurlyeq^{x_{t+1}}_{k-E} (k_2,P),\tag{3.3}\end{equation}
where $x_1,x_2,\cdots,x_t,x_{t+1}\in\{a,b,c\}$.

\if false Note that, the condition ``$\#\{j: R_j\supseteq Q_{i_j}\ \mbox{\rm for som } Q_{i_j}\}\geq k_1$" in (a$'$) and (b$'$) seems strange but is necessary and reasonable.  This is because that $(k_{1}, Q)\preccurlyeq^{x'} (k_2, P)$, $x\in\{a,b,c\}$ should require not only $Q\preccurlyeq^{x'} P$, but also that every $k_2$-partition $R=R_1|R_2|\ldots|R_{k_2}$ of $P$ is a refinement of some $k_1$-partition $S=S_1|S_2|\ldots|S_{k_1}$. This is satisfied automatically when $Q\preccurlyeq^c P$ and $2\leq k_1\leq k_2\leq m$. Here we say that $R=R_1|R_2|\ldots|R_{k_2}$ is a refinement of $S=S_1|S_2|\ldots|S_{k_1}$ if $S\preccurlyeq R$. \if false $k_1\leq k$ and every $S_i$ is a union of some $R_{1_i}\bigcap(\bigcup_{s=1}^r Q_s)$,$R_{2_i}\bigcap(\bigcup_{s=1}^r Q_s)$, $\ldots$, $R_{j_i}\bigcap(\bigcup_{s=1}^r Q_s)$, that is, $S_i=(\bigcup_{t=1_i}^{j_i} R_t)\bigcap (\bigcup_{s=1}^r Q_s)$.\fi
For instance, if  $Q\preccurlyeq^a P$ and $(k_1,k_2)\in\{(r-1,m-1), (r,m)\}$, then we must have  $(k_1,Q)\preccurlyeq^{a'}(k,P)$.\fi

Thus  the hierarchy condition for a $k$-entanglement measure $E^{(k,n)}$ can be stated as follows.

{\bf ($k$-EM4)} (Hierarchy condition) $\{E^{(s,m)}: 2\leq s\leq \min\{k,m\}, m=2,3,\cdots, n\}$ has  the following three properties: for any $\rho=\rho_{1,2,\cdots,n}\in{\mathcal S}(H_1\otimes H_2\otimes\cdots\otimes H_n)$ and any $P=P_1|P_2|\cdots|P_m$, $Q=Q_1|Q_2|\cdots |Q_r\in\mathcal{SP}_{n}$, regarding $\rho_P={\rm Tr}_{P^c}(\rho)\in{\mathcal S}(H_{P_1}\otimes H_{P_2}\otimes\cdots\otimes H_{P_m})$ as  $m$-partite state and $\rho_Q={\rm Tr}_{Q^c}(\rho)\in{\mathcal S}(H_{Q_1}\otimes H_{Q_2}\otimes\cdots\otimes H_{Q_r})$ as  $r$-partite state,

($k$-EM4a) $(k_{1}, Q)\preccurlyeq^{a}_{k-E} (k_2, P)$ implies $E^{(k_1,r)}(\rho_Q, Q)\leq E^{(k_2,m)}(\rho_P, P)$;

($k$-EM4b) $(k_{1}, Q)\preccurlyeq^{b}_{k-E} (k_2, P)$ implies $E^{(k_1,r)}(\rho_Q, Q)\leq E^{(k_2,m)}(\rho_P, P)$;

($k$-EM4c) $(k_{1}, Q)\preccurlyeq^{c}_{k-E} (k_2, P)$ implies $E^{(k_1,m)}(\rho_Q, Q)\leq E^{(k_2,m)}(\rho_P, P)$.

Thus $E^{(k,n)}$ is a true $k$-entanglement measure if it satisfies the conditions ($k$-EM1) - ($k$-EM5). It is clear that, if $k=n$, the true $n$-entanglement measure $E^{(n)}=E^{(n,n)}$ is exactly a true entanglement measure discussed in Example 3.1, because in this situation, we must have $k_1=r$, $k_2=m$, and $(r, Q)\preccurlyeq_{k-E} (m, P)$ if and only if $Q\preccurlyeq P$.

 In \cite{GYHQH}, an EW-based true $k$-entanglement measure $E_{w}^{(k,n)}$ without convex roof extension was proposed, which is defined by
\begin{equation}
E_{w}^{(k,n)}(\rho)=\mathop{\rm sup}\limits_{L\in\mathcal{B}_{1}^{+}}
\{{\rm max}\{{\rm Tr}(L\rho)-\lambda,0\}:g_{n}^{(k)}(L)\leq\lambda\leq\|L\|\}
\tag{3.4}
\end{equation}
for any $\rho\in {\mathcal S}(H_{1}\otimes H_{2}\otimes\cdots\otimes H_{n})$, where $\mathcal{B}_{1}^{+}=\{L\in\mathcal{B}(H_{1}\otimes H_{2}\otimes\cdots\otimes H_{n}):L\geq0,\|L\|\leq1\}$ and $g_{n}^{(k)}(L)={\rm max}\{\langle\phi|L|\phi\rangle:|\phi\rangle\in H_{1}\otimes H_{2}\otimes\cdots\otimes H_{n} \ \mbox{\rm is } k\mbox{\rm -separable pure vector state} \}$. In addition,  it was shown in \cite{GYHQH} that $E_{w}^{(k,n)}$   satisfies the conditions ($k$-EM1)-($k$-EM5), making it  a true $k$-entanglement measure. Therefore, $k$-entanglement is a completely symmetric multipartite quantum resource. Moreover, $E_{w}^{(k,n)}$ has other nice properties such as convexity, subadditivity and computability.

{\bf Example 3.3.}  The $k$-partite entanglement is a completely symmetric multipartite quantum resource.

Based on the depth of multipartite entanglement and the technical difficulty in preparing multipartite entangled states, an $n$-partite quantum state can be categorized as $k$-producible states $(1\leq k \leq n-1)$ and $(k+1)$-partite entangled states.  Recall that, for $1\leq k\leq n-1$,  a pure state $|\psi\rangle\in H_1\otimes H_2\otimes\cdots\otimes H_n $ is said to be $k$-producible if there exists a partition $\{P_1,P_2,\cdots, P_m\}$ of $\{1,2,\cdots, n\}$ such that $|\psi\rangle$ is a product vector of the composite system $H_{P_1}\otimes H_{P_2}\otimes\cdots\otimes H_{P_m}$, where $\# P_j$, the number of particles in subset \( P_j \), is not greater than \( k \) for each $j=1,2,\cdots, m$, that is, the depth of $P$ is not greater than $k$. A state $\rho\in {\mathcal S}(H_1\otimes H_2\otimes\cdots\otimes H_n )$ is said to be $k$-producible if it is a convex combination of $k$-producible pure states or the trace-norm limit of such convex combinations. Otherwise, we say that \( \rho \) contains \( (k+1) \)-partite entanglement or $\rho$ is $(k+1)$-partite entangled \cite{OGH}.  The set \(\mathcal D_k= \mathcal D_k(H_1\otimes H_2\otimes\cdots\otimes H_n) \) of all \( k \)-producible quantum states
 is a closed convex subset of ${\mathcal S}(H_1\otimes H_2\otimes\cdots\otimes H_n )$ and $D_k\subset D_l$ if $k<l$.   Thus, the set of  entangled states is classified into $k$-partite entangled states, $k=n-1,\cdots, 3,2,1$. $k$-partite entanglement is also a completely symmetric MQC.

\if false  Then, with $(k-1)$-producible states as free states and LOCCs as free operations, is the $k$-partite entanglement a multipartite quantum resource?\fi

 Most recently, Ref.\cite{QUA} affirmatively answered this question by proposing a true $k$-partite entanglement measure $\mathbb E_w^{(k,n)}$ that does not require  convex roof extension, satisfying the faithfulness ($k$-PEM1), non-increasing trend under LOCC ($k$-PEM2), unification condition ($k$-PEM3), hierarchy condition ($k$-PEM4), and  symmetry ($k$-PEM5).

 Similar to $k$-entanglement, the unification condition and the hierarchy condition of $k$-partite entanglement measure are relatively complex. The unification for a $k$-partite entanglement measure $\mathbb{E}^{(k,n)}$ reads as follows.

{\bf ($k$-PEM3)} (Unification condition)
All $\mathbb{E}^{(s,m)}$s are defined in the same way as $\mathbb{E}^{(k,n)}$  so that for any $P=P_1|P_2|\ldots|P_m\in\mathcal{SP}_n$ and $2\leq s\leq \min\{k,m\}$, the $s$-partite entanglement $\mathbb E^{(s,m)}(\rho_P,P)$ in $\rho_P\in\mathcal S(H_{P_1}\otimes H_{P_2}\otimes\cdots\otimes H_{P_m})$ can be measured for any state $\rho\in S(H_{1}\otimes H_{2}\otimes\cdots\otimes H_{n})$ without causing any confusions.
\if false
and $\{\mathbb{E}^{(s,m)}: 2\leq s\leq \min\{k,m\}, 2\leq m\leq n\}$ is gets along well  with each other.\fi

The hierarchy condition for $k$-partite entanglement measure ensures that, if the hierarchy of a sub-repartition $Q=Q_1|Q_2|\cdots|Q_r$ with $2\leq k_1\leq r$ is lower than the hierarchy of a sub-repartition $P=P_1|P_2|\cdots|P_m$ with $2\leq k_2\leq m$ in terms of $k$-partite entanglement, denoted by $(k_1,Q)\preccurlyeq_{k-PE} (k_2,P)$, then for any state $\rho\in \mathcal S(H_1\otimes H_2\otimes\cdots\otimes H_n)$,  the  $k_{1}$-partite entanglement of its reduced state $\rho_Q$ with respect to the part system $H_Q=H_{Q_1}\otimes H_{Q_2}\otimes\cdots\otimes H_{Q_r}$ never exceeds the  $k_{2}$-partite entanglement of its reduced state $\rho_P$ with respect to the part system $H_P=H_{P_1}\otimes H_{P_2}\otimes\cdots\otimes H_{P_m}$. Thus, to clarify the conception of the hierarchy condition for $k$-partite entanglement measure, we should first give the exact meaning of $(k_1,Q)\preccurlyeq_{k-PE} (k_2,P)$. If $(k_1,Q)\preccurlyeq_{k-PE} (k_2,P)$, we also say that $(k_1,Q)$ is coarser than $(k_2,P)$ with respect to $k$-partite entanglement. Note that, by the definition of $k$-partite entanglement, if $k_1<k$, the $k$-partite entanglement contained in a state $\rho$ should not exceed  the $k_1$-partite entanglement contained in $\rho$.

By \cite{QUA}, for any $P=P_1|P_2|\cdots|P_m$, $Q=Q_1|Q_2|\cdots|Q_r\in\mathcal{SP}_{n}$,  there are three cases in which $(k_1,Q)$ is coarser than $(k_2,P)$ with respect to $k$-partite entanglement:

{\bf ($k$-PEa)} If $2\leq k_2\leq k_1\leq r$,  $ k_{2}\leq m$ and $Q\preccurlyeq^a P$, we say that the hierarchy of $(k_1,Q)$ is lower than the hierarchy of $(k_2,P)$ in type (a), denoted by $(k_1,Q)\preccurlyeq^{a}_{k-PE} (k_2, P)$;

{\bf ($k$-PEb)} If $2\leq k_2\leq k_1\leq r$,  $ k_{2}\leq m$, $Q\preccurlyeq^b P$, and for any partition $R=R_1|R_2|\cdots|R_{t}$ $(t\leq m)$ of $P$ with depth at most $k_{2}-1$ and any $i=1,2,\cdots, r$, there exists $R_{i_j}$ such that $Q_{i}\cap R_{i_j}\neq\emptyset$ and  $\#\{s: Q_{s}\cap (\cup_{j} R_{i_j})\neq \emptyset\}\leq k_1-1$, we say that the hierarchy of $(k_1,Q)$ is lower than the hierarchy of $(k_2,P)$ in type (b), denoted by $(k_{1}, Q)\preccurlyeq^{b}_{k-PE} (k_2, P)$;

{\bf ($k$-PEc)} If $r=m$, $2\leq k_2\leq k_1\leq m$  and $Q\preccurlyeq^c P$, we say that the hierarchy of $(k_1,Q)$ is lower than the hierarchy of $(k_2,P)$ in type (c), denoted by $(k_{1}, Q)\preccurlyeq^{c}_{k-PE} (k_2, P)$.

Thus, for any $P=P_1|P_2|\cdots|P_m\in\mathcal{SP}_{n}$ and $Q=Q_1|Q_2|\ldots |Q_r\in\mathcal{SP}_{n}$, we say that the hierarchy of $(k_{1}, Q)$ is lower than the hierarchy of   $(k_2, P)$ with respect to $k$-partite entanglement, or, in other words, $(k_{1}, Q)$ is coarser than  $(k_2, P)$ with respect to $k$-partite entanglement, denote by $(k_{1}, Q)\preccurlyeq_{k-PE} (k_2, P)$,   if   there are some sub-repartitions $R_1,R_2,\cdots, R_t\in\mathcal{SP}_{n}$ and positive integers $r_1, r_2,\cdots, r_t$ such that
\begin{equation}(k_1,Q)\preccurlyeq_{k-PE}^{x_1} (r_1,R_1)\preccurlyeq_{k-PE}^{x_2} (r_2,R_2) \preccurlyeq_{k-PE}^{x_3}\cdots \preccurlyeq_{k-PE}^{x_t} (r_t,R_t)\preccurlyeq_{k-PE}^{x_{t+1}} (k_2,P),\tag{3.5}\end{equation}
where $x_1,x_2,\cdots,x_t,x_{t+1}\in\{a,b,c\}$.

\if false Note that, the condition ``$\#\{j: R_j\supseteq Q_{i_j}\ \mbox{\rm for som } Q_{i_j}\}\geq k_1$" in (a$'$) and (b$'$) seems strange but is necessary and reasonable.  This is because that $(k_{1}, Q)\preccurlyeq^{x'} (k_2, P)$, $x\in\{a,b,c\}$ should require not only $Q\preccurlyeq^{x'} P$, but also that every $k_2$-partition $R=R_1|R_2|\ldots|R_{k_2}$ of $P$ is a refinement of some $k_1$-partition $S=S_1|S_2|\cdots|S_{k_1}$. This is satisfied automatically when $Q\preccurlyeq^c P$ and $2\leq k_1\leq k_2\leq m$. Here we say that $R=R_1|R_2|\ldots|R_{k_2}$ is a refinement of $S=S_1|S_2|\cdots|S_{k_1}$ if $S\preccurlyeq R$. \if false $k_1\leq k$ and every $S_i$ is a union of some $R_{1_i}\bigcap(\bigcup_{s=1}^r Q_s)$,$R_{2_i}\bigcap(\bigcup_{s=1}^r Q_s)$, $\ldots$, $R_{j_i}\bigcap(\bigcup_{s=1}^r Q_s)$, that is, $S_i=(\bigcup_{t=1_i}^{j_i} R_t)\bigcap (\bigcup_{s=1}^r Q_s)$.\fi
For instance, if  $Q\preccurlyeq^a P$ and $(k_1,k_2)\in\{(r-1,m-1), (r,m)\}$, then we must have  $(k_1,Q)\preccurlyeq^{a'}(k,P)$.\fi

Thus  the hierarchy condition for a $k$-partite entanglement measure $\mathbb{E}^{(k,n)}$ can be stated as follows.

{\bf ($k$-PEM4)} (Hierarchy condition) $\{\mathbb{E}^{(s,m)}: 2\leq s\leq \min\{k,m\}, m=2,3,\cdots, n\}$ has  the following three properties: for any $\rho=\rho_{1,2,\cdots,n}\in{\mathcal S}(H_1\otimes H_2\otimes\cdots\otimes H_n)$ and any $P=P_1|P_2|\cdots|P_m$, $Q=Q_1|Q_2|\cdots |Q_r\in\mathcal{SP}_{n}$, regarding $\rho_P={\rm Tr}_{P^c}(\rho)\in{\mathcal S}(H_{P_1}\otimes H_{P_2}\otimes\cdots\otimes H_{P_m})$ as  $m$-partite state and $\rho_Q={\rm Tr}_{Q^c}(\rho)\in{\mathcal S}(H_{Q_1}\otimes H_{Q_2}\otimes\cdots\otimes H_{Q_r})$ as  $r$-partite state,

($k$-PEM4a) $(k_1,Q)\preccurlyeq^{a}_{k-PE} (k_2, P)$  implies $\mathbb{E}^{(k_1,r)}(\rho_Q, Q)\leq \mathbb{E}^{(k_2,m)}(\rho_P, P)$;

($k$-PEM4b) $(k_1,Q)\preccurlyeq^{b}_{k-PE} (k_2, P)$ implies $\mathbb{E}^{(k_1,r)}(\rho_Q, Q)\leq \mathbb{E}^{(k_2,m)}(\rho_P, P)$;

($k$-PEM4c) $(k_1,Q)\preccurlyeq^{c}_{k-PE} (k_2, P)$ implies $\mathbb{E}^{(k_1,m)}(\rho_Q, Q)\leq \mathbb{E}^{(k_2,m)}(\rho_P, P)$.

We call that $\mathbb{E}^{(k,n)}$ is a true $k$-partite entanglement measure if it satisfies the conditions ($k$-PEM1) - ($k$-PEM5).

 The hierarchy condition for $k$-partite entanglement measures was also discussed in \cite{JYF}, with the conditions ($k$-PEM4) without $\#\{s: Q_{s}\cap (\cup_{j} R_{i_j})\neq \emptyset\}\leq k_1-1$ in the definition of $(k_{1}, Q)\preccurlyeq_{k-PE}^{b} (k_2, P)$. Moreover,  a quantification of $k$-partite entanglement  from the minimal sum was proposed in \cite{JYF}, that is,
\begin{equation}\mathbb{E}^{(k)}(|\psi\rangle)=\frac{1}{2}\mathop{\rm min}\limits_{X\in{\mathbb P}_n^{\leq k-1}}\sum_{t=1}^{m}h(\rho_{X_{t}}) \tag{3.6}\end{equation}
for any $|\psi\rangle\in H_1\otimes H_2\otimes\cdots\otimes H_n$, where $h$ is a non-negative concave function and ${\mathbb P}_n^{\leq k-1}$ is the set of all partitions $X=X_{1}|X_{2}|\cdots|X_{m}$ of $\{1,2,\cdots,n\}$  with the depth of $X$ at most $k-1$. $\rho_{X_{t}}={\rm Tr}_{X_{t}^{c}}(|\psi\rangle\langle\psi|)$ and $X_{t}^{c}$ denotes the subsystems complementary to those of $X_{t}$. Then the value for mixed state is defined via  the convex-roof extension. In \cite{JYF}, it was shown that $\mathbb{E}^{(k)}$   satisfies the conditions ($k$-PEM1) - ($k$-PEM3), ($k$-PEM4a), ($k$-PEM4c) and ($k$-PEM5).  However, for condition ($k$-PEM4b), in the absence of  condition $\#\{s: Q_{s}\cap (\cup_{j} R_{i_j})\neq \emptyset\}\leq k_1-1$, $\mathbb{E}^{(k)}$ only satisfies the requirements when $k_{1}=k_{2}=2$ and $h$ is subadditive.

 Now we prove that if $h$ is subadditive, then $\mathbb{E}^{(k)}$ is a true $k$-partite entanglement measure. In fact, we only need to check the following proposition.

{\bf Proposition 3.1.} If $h$ is subadditive, then $\mathbb{E}^{(k)}$ satisfies the property ($k$-PEM4b).

{\bf Proof.} It suffices to demonstrate that condition ($k$-PEM4b) is satisfied for any pure state $|\psi\rangle\in H_1\otimes H_2\otimes\cdots\otimes H_n$.

Without loss of generality, we may assume that $P=P_1|P_2|\cdots|P_m$ is an $m$-partition of $\{1,2,\cdots,n\}$. Then $\rho_Q=\rho_P=\rho$ for any $\rho=|\psi\rangle\langle\psi|\in{\mathcal S}(H_1\otimes H_2\otimes\cdots\otimes H_n)$.

Denote the depth of $P$ by
depth$(P)=\max\{\#P_j: j=1,2,\cdots ,m\}$. Let $\mathbb{P}_{m}^{\leq k_{2}-1, P}$ be the set of all partitions of $P=P_1|P_2|\cdots|P_m$  with depth at most $k_{2}-1$ related to $P$, that is, $R=R_1|R_2|\cdots|R_{t}\in \mathbb{P}_{m}^{\leq k_{2}-1, P}$ if and only if $R$ is a partition of $P$ and every $R_j$ is a union of at most $k_2-1$ $P_{j_i}$s. Denote by depth$^P(R)=\max_j\{ \#\{{j_i}: R_j=\cup P_{j_i}\}$, the depth of $R$ related to $P$. Then
\begin{equation}\begin{array}{rl}
\mathbb{P}_{m}^{\leq k_{2}-1, P}=\{R=R_1|R_2|\cdots|R_{t}:& R_{j} \ \mbox{\rm is a union of some } P_{j_i}s \ \mbox{\rm for each }  \\& j=1,2, \cdots, t \ \   \mbox{\rm and depth}^P(R)\leq k_{2}-1 \}.\end{array}\tag{3.7}\end{equation}
Similarly, let $\mathbb{P}_{r}^{\leq k_{1}-1, Q}$ be the set of all partitions of $Q=Q_1|Q_2|\cdots|Q_r$  with depth related to $Q$ at most $k_{1}-1$. \if false that is,
$$\begin{array}{rl}
\mathbb{P}_{r}^{\leq k_{1}-1, Q}=\{S=S_1|S_2|\cdots|S_{t_{1}}:& S_{j} \ \mbox{\rm is a union of some } Q_{i} \ \mbox{\rm for each }\\ &   j=1,2, \cdots, t_{1} \mbox{\rm and depth}(S)\leq k_{1}-1 \}.\end{array}$$\fi

Clearly, for any partition $S=S_1|S_2|\cdots|S_{t_{1}}\in\mathbb{P}_{r}^{\leq k_{1}-1, Q}$, as $k_{2}\leq k_{1}\leq r$ and $Q\preccurlyeq^b P$, there exists a partition $R=R_1|R_2|\cdots|R_{t_{2}}\in\mathbb{P}_{m}^{\leq k_{2}-1, P}$ such that  each $S_{i}$ is a union of some $R_{j}$, in this case, we say that $R\in \mathbb{P}_{m}^{\leq k_{2}-1, P}$ is a  refinement of $S\in \mathbb{P}_{r}^{\leq k_{1}-1, Q}$. The set of all refinements of $S$ is denoted as $\mathcal{R}_{S}$; then $R\in\mathcal{R}_{S}$. Conversely, for any partition $R=R_1|R_2|\cdots|R_{t_{2}}\in\mathbb{P}_{m}^{\leq k_{2}-1, P}$, as for any $i=1,2,\cdots, r$,  $\#\{s: Q_{s}\cap (\cup_{j} R_{i_j})\neq \emptyset\}\leq k_1-1$ (where $R_{i_j}$ satisfies $Q_{i}\cap R_{i_j}\neq\emptyset$),  there must exist some $S=S_1|S_2|\cdots|S_{t_{1}}\in\mathbb{P}_{r}^{\leq k_{1}-1, Q}$ such that  each $S_{i}$ is a union of some $R_{j}$, that is, $R\in\mathcal{R}_{S}$.

Since $h$ is subadditive,  for any $S=S_1|S_2|\cdots|S_{t_1}\in\mathbb{P}_{r}^{\leq k_{1}-1, Q}$, and its refinement $R=R_1|R_2|\cdots|R_{t_{2}}\in\mathbb{P}_{m}^{\leq k_{2}-1, P}$,  we see that, $\sum_{i=1}^{t_{1}}h(\rho_{S_{i}})\leq\sum_{j=1}^{t_{2}}h(\rho_{R_{j}})$.

Therefore, for any $\rho=|\psi\rangle\langle\psi|\in{\mathcal S}(H_1\otimes H_2\otimes\cdots\otimes H_n)$,
$$\begin{array}{rl}
 \mathbb{E}^{(k)}(\rho,Q)=& \mathop{\rm min}\limits_{S\in\mathbb{P}_{r}^{\leq k_{1}-1, Q}} \frac{1}{2}\sum_{i=1}^{t_{1}}h(\rho_{S_{i}})\\
 \leq & \mathop{\rm min}\limits_{R\in\mathbb{P}_{m}^{\leq k_{2}-1, P}\cap\mathcal{R}_{S}} \frac{1}{2}\sum_{j=1}^{t_{2}}h(\rho_{R_{j}})\\
 =&\mathop{\rm min}\limits_{R\in\mathbb{P}_{m}^{\leq k_{2}-1, P}} \frac{1}{2}\sum_{j=1}^{t_{2}}h(\rho_{R_{j}}) \\
 =&\mathbb{E}^{(k)}(\rho,P)
\end{array}$$
as desired.\hfill$\Box$

Consequently, our hierarchy condition ($k$-PEM4) for $k$-partite entanglement measures is more reasonable than the hierarchy condition presented in  \cite{JYF}, and if $h$ is subadditive, $\mathbb{E}^{(k)}$ defined by Eq.(3.6)  is a true $k$-partite entanglement measure in our axiomatic framework.

Furthermore, in \cite{QUA}, a true $k$-partite entanglement measure $\mathbb{E}_{w}^{(k,n)}$ without convex roof extension was proposed, which is defined by
\begin{equation}
\mathbb{E}_{w}^{(k,n)}(\rho)=\sup_{L\in  \mathcal{B}_1^+} \max\{ 0, {\rm Tr}(L\rho)-f^{(k,n)}_{\max}(L)\}
\tag{3.8}
\end{equation}
for any $\rho\in {\mathcal S}(H_{1}\otimes H_{2}\otimes\cdots\otimes H_{n})$, where $\mathcal{B}_{1}^{+}=\{L\in\mathcal{B}(H_{1}\otimes H_{2}\otimes\cdots\otimes H_{n}): 0\leq L\leq I\}$ and $f^{(k,n)}_{\max}(L)=\sup\{ {\rm Tr}(L\sigma): \sigma\ \mbox{\rm is } (k-1)\mbox{\rm -producible pure vector state}\}$. In \cite{QUA}, it was shown that $\mathbb{E}_{w}^{(k,n)}$   satisfies the conditions ($k$-PEM1)-($k$-PEM5) and enjoys the convexity and subadditivity, making it a true $k$-partite entanglement measure.

Therefore, $k$-partite entanglement is a completely symmetric multipartite quantum resource.

\subsection{Completely symmetric multipartite quantum resources beyond entanglement}

In addition to multipartite entanglement, $k$-entanglement, and $k$-partite entanglement, there are other kinds of completely symmetric MQCs that are also useful multipartite quantum resources.

{\bf Example 3.4.} The non-MPPT  is a completely symmetric multipartite quantum resource.

From \cite{RAE}, a multipartite state is called a multipartite positive partial transpose (MPPT) state when it is PPT with respect to all bipartitions; otherwise, it is a non-MPPT state.  The non-MPPT is a completely symmetric MQC.

For an $n$-partite system $H=H_1\otimes H_2\otimes\cdots\otimes H_n$, denote by $\mathcal P_n^{(2)} $ the set of all bipartitions of $\{1,2,\cdots,n\}$.  The set of all MPPT states is
$$\mathcal{MP}=\{\rho\in\mathcal{S}(H_{1}\otimes H_{2}\otimes\cdot\cdot\cdot\otimes H_{n}):\forall \ P=P_1|P_2\in\mathcal P_n^{(2)},\rho^{ {\rm T}_{1}}\geq0\},$$
which is a convex closed subset in ${\mathcal S}(H)$, where ${\rm T}_{1}$ is the partial transpose with respect to the subsystem $H_{P_{1}}$.

\if false A quantum channel is a MPPT channel if it sends every MPPT state to a MPPT state.\fi As all bipartitions of the system are involved, it is natural to consider LOCCs as free operations.  It is clear that every LOCC sends MPPT states to MPPT states.
In fact, if $\Phi$ is LOCC, then $\Phi$ has an operator-sum representation $\Phi(\rho)=\sum_j K_j\rho K_j^\dag$ with all the Kraus operators $K_j$ have the product form $K_j=K_{1j}\otimes K_{2j}\otimes\cdots\otimes K_{nj}$. For any bipartition $P=P_1|P_2\in\mathcal P_n^{(2)}$, $K_j=E_j\otimes F_j$ with respect to $H=H_{P_1}\otimes H_{P_2}$ and $$\Phi(\rho)^{\rm T_1}=\sum_{j} (\bar{E_j}\otimes F_j)\rho^{\rm T_1}(\bar{E_j}\otimes F_j)^\dag,$$
where $\bar{A}=(\bar{a}_{ij})$ if $A =(a_{ij})$. Hence, $\rho$ is PPT with respect to $P=P_1|P_2$ implies that $\Phi(\rho)$ is PPT.

 With all MPPT states being the free states and  all LOCCs being the free operations, we claim that non-MPPT is a  completely symmetric multipartite quantum resource.

 To check this, we need  propose a true non-MPPT measure $p^{(n)}$ as follows.

\if false The free operations are all full LOCC operations $\Lambda$, that is,  all trace-preserving, completely positive operations $\Lambda(\rho)=\sum_{i}K_{i}\rho K_{i}^{\dag}$ with $\sum_{i}K_{i}^{\dag}K_{i}=I$, and each operator $K_{i}=A_{i}\otimes B_{i}\otimes\cdot\cdot\cdot\otimes F_{i}$ has a tensor product form.  Now, we introduce a measure specifically
tailored for non-MPPT. \fi

Let
$$\mathcal{DW}^{(n)}=\{W : \ \|W\|\leq 1, \ \forall P=P_1|P_2\in\mathcal P_n^{(2)}, \ \exists\ 0\leq M_{P},N_{P}\  \mbox{\rm such that}\  W=M_{P}+N_{P}^{\rm T_1}\}$$
be the set of  fully decomposable witnesses.
For any multipartite quantum state $\rho\in\mathcal{S}(H_{1}\otimes H_{2}\otimes\cdot\cdot\cdot\otimes H_{n})$, define
\begin{equation}p^{(n)}(\rho)=\max_{W\in\mathcal{DW}^{(n)}} |\min\{{\rm Tr}(\rho W),0\}|.\tag{3.9}\end{equation}
If $\rho\in \mathcal{MP}$, then for any bipartition $P=P_1|P_2\in\mathcal P_n^{(2)}$, $\rho^{\rm T_1}\geq 0$. This implies that, for any $W\in{\mathcal{DW}}^{(n)}$, writing $W=M_P+N_P^{\rm T_1}$, ${\rm Tr}(W\rho)\geq0$ and hence $p^{(n)}(\rho)=0$. The converse is also true because if $\rho\notin \mathcal{MP}$, then, there is a bipartition $P=P_1|P_2\in\mathcal P_n^{(2)}$ such that $\rho$ is not PPT with respect to $H_{P_1}\otimes H_{P_2}$, and thus, by  \cite{BTO}, there exists a $W\in \mathcal{DW}^{(n)}$ such that ${\rm Tr}(W\rho)<0$, which leads to $p^{(n)}(\rho)>0$.  So $p^{(n)}(\rho)=0$ if and only if $\rho\in \mathcal{MP}$. Furthermore,  $p^{(n)}$ is non-increasing under LOCCs  \cite{BTO}, that is, $p^{(n)}$ satisfies conditions (MQCM1)-(MQCM2). And, it is easy to see that $p^{(n)}$  satisfies conditions (MQCM3) and (MQCM5). With the hierarchy relation $\preccurlyeq$ between sub-repartitions defined in Eq.(3.1), we will conduct a simple verification that $p^{(n)}$ satisfies the condition (MQCM4) which can be restated as

{\bf (CS-MQCM4)} (Hierarchy condition) As a quantification  of non-MPPT, $\mathcal C^{(n)}$ satisfies the following three conditions: for any $P=P_1|P_2|\cdots|P_m$, $Q=Q_1|Q_2|\cdots|Q_r\in\mathcal{SP}_{n}$ and any $\rho\in {\mathcal S}(H_1\otimes H_2\otimes \cdots\otimes H_n)$,

(CS-MQCM4a) $Q\preccurlyeq^a P$ implies $\mathcal C^{(r)}(\rho_Q, Q)\leq \mathcal C^{(m)}(\rho_P, P)$;

(CS-MQCM4b) $Q\preccurlyeq^b P$ implies $\mathcal C^{(r)}(\rho_Q, Q)\leq \mathcal C^{(m)}(\rho_P, P)$;

(CS-MQCM4c) $Q\preccurlyeq^c P$ implies $\mathcal C^{(m)}(\rho_Q, Q)\leq \mathcal C^{(m)}(\rho_P, P)$.

We prove $p^{(n)}$ meets (CM-MQCM4) by three propositions. For any $P=P_1|P_2|\ldots|P_m\in\mathcal{SP}_n$, denote $$\begin{array}{rl} \mathcal{DW}_P^{(m)}=& \{W\in\mathcal B(H_P): \|W\|\leq 1\ \ {\rm and}\ \forall R=R_1|R_2\in{\mathcal P}^{(2,P)},\\ & \ \ \exists \ 0\leq M_R,N_R   {\rm such \ that }\ W=M_R+N_R^{T_1}\},
\end{array}$$
where ${\mathcal P}^{(2,P)}$ is the set of all bipartitions of $H_P=H_{P_1}\otimes H_{P_2}\otimes\cdots\otimes H_{P_m}$.

{\bf Proposition 3.2.} For any $\rho\in\mathcal S(H_1\otimes H_2\otimes \cdots\otimes H_n)$ and sub-repartitions $P=P_1|P_2|\cdots|P_m,\ Q=Q_1|Q_2|\cdots |Q_r\in{\mathcal SP}_n$, $Q\preccurlyeq^a P$ implies $p^{(r)}(\rho_Q,Q)\leq p^{(m)}(\rho_P,P)$.

{\bf Proof.} Without loss of generality, it suffices to show that, for any $\rho_{1,2,\cdots,n}\in{\mathcal S}(H_1\otimes H_2\otimes\cdots\otimes H_n)$ and a subset $\{1,2,\cdots, r\}\subseteq\{1,2,\cdots,n\}$, we have
$$p^{(r)}(\rho_{1,2,\cdots, r})\leq p^{(n)}(\rho_{1,2,\cdots,n}),$$
 where $\rho_{1,2,\cdots,r}={\rm Tr}_{\{r+1,r+2,\cdots,n\}}(\rho_{1,2,\cdots,n})$ is the reduced state of $\rho_{1,2,\cdots,n}$ to the subsystem $\{1,2,\cdots,r\}$.
In fact,
$$\begin{array}{rl}
p^{(r)}(\rho_{1,2,\cdots,r})=&\mathop{\rm max}\limits_{W\in\mathcal{DW}_Q^{(r)}} |\min\{{\rm Tr}(\rho_{1,2,\cdots, r} W),0\}|\nonumber\\=& \mathop{\rm max}\limits_{W\otimes I_{Q^c}\in\mathcal{DW}^{(n)}} |\min\{{\rm Tr}[\rho_{1,\cdots, n} (W\otimes I_{Q^c})],0\}|\nonumber\\ \leq&\mathop{\rm max}\limits_{W\in\mathcal{DW}^{(n)}} |\min\{{\rm Tr}(\rho_{1,\cdots, n} W),0\}|\nonumber\\ =& p^{(n)}(\rho_{1,2,\cdots,n})
\nonumber
\end{array}$$
as desired. \hfill$\Box$

{\bf Proposition 3.3.} For any $\rho\in\mathcal S(H_1\otimes H_2\otimes \cdots\otimes H_n)$ and sub-repartitions $P=P_1|P_2|\cdots|P_m,\ Q=Q_1|Q_2|\cdots |Q_r\in{\mathcal SP}_n$, $Q\preccurlyeq^b P$ implies $p^{(r)}(\rho_Q,Q)\leq p^{(m)}(\rho_P,P)$.

{\bf Proof.}  It suffices to show that,  for any $r$-partition $Q=Q_1|Q_2|\cdots |Q_r$ of $\{1,2,\cdots,n\}$ and any $\rho=\rho_{1,2,\cdots,n}\in {\mathcal S}(H_{1}\otimes H_{2}\otimes\cdots\otimes H_{n})$,  regarding $\rho$ as a $r$-partite state $\rho\in {\mathcal S}(H_{Q_1}\otimes H_{Q_2}\otimes\cdots\otimes H_{Q_r})$, we have
$$p^{(r)}(\rho,Q)\leq p^{(n)}(\rho).$$

  We first observe that every bipartition of $Q$  is also a   bipartition of $\{1,2,\cdots,n\}$, so we have $\mathcal{DW}_{Q}^{(r)}\subset \mathcal{DW}^{(n)}$. It follows that
$$\begin{array}{rl}
p^{(r)}(\rho,Q)=&\mathop{\rm max}\limits_{W\in\mathcal{DW}_{Q}^{(r)}} |\min\{{\rm Tr}(\rho W),0\}|\nonumber\\ \leq&\mathop{\rm max}\limits_{W\in\mathcal{DW}^{(n)}} |\min\{{\rm Tr}(\rho W),0\}|\nonumber\\ =& p^{(n)}(\rho)
\nonumber
\end{array}$$
completing the proof. \hfill$\Box$

{\bf Proposition 3.4.} For any $\rho\in\mathcal S(H_1\otimes H_2\otimes \cdots\otimes H_n)$ and sub-repartitions $P=P_1|P_2|\cdots|P_m,\ Q=Q_1|Q_2|\cdots |Q_r\in{\mathcal SP}_n$, $Q\preccurlyeq^c P$ implies $p^{(m)}(\rho_Q,Q)\leq p^{(m)}(\rho_P,P)$ as $r=m$.

{\bf Proof.}   Without loss of generality, let $P=P_1|P_2|\cdots |P_m$ be an $m$-partition of $\{1,2,\cdots,n\}$ and $Q=Q_1|Q_2|\cdots |Q_m$ such that $Q_j\subseteq P_j$.  Then, for any $\rho=\rho_{1,2,\cdots,n}\in{\mathcal S}(H_1\otimes H_2\otimes\cdots\otimes H_n)$,   regarding also $\rho\in{\mathcal S}(H_{P_1}\otimes H_{P_2}\otimes\cdots\otimes H_{P_m})$ and $\rho_Q={\rm Tr}_{Q^c}(\rho)\in{\mathcal S}(H_{Q_1}\otimes H_{Q_2}\otimes\cdots\otimes H_{Q_m})$  as   $m$-partite states with respect to $P$ and $Q$, respectively, we have
$$\begin{array}{rl}
p^{(m)}(\rho_Q,Q)=&\mathop{\rm max}\limits_{W\in\mathcal{DW}_{Q}^{(m)}} |\min\{{\rm Tr}(\rho_{Q} W),0\}|\nonumber\\=&\mathop{\rm max}\limits_{W\otimes I_{Q^{c}}\in\mathcal{DW}_{P}^{(m)}} |\min\{{\rm Tr}[\rho (W\otimes I_{Q^c})],0\}|\nonumber\\ \leq&\mathop{\rm max}\limits_{W\in\mathcal{DW}_{P}^{(m)}} |\min\{{\rm Tr}(\rho W),0\}|\nonumber\\ =& p^{(m)}(\rho,P)
\nonumber
\end{array}$$
as desired. \hfill$\Box$

By Propositions 3.2-3.4,  $p^{(n)}$ satisfies conditions (MQCM1)-(MQCM5) with hierarchy relation $\preccurlyeq$, making it a true measure of   non-MPPT. Consequently, the  non-MPPT  is a completely symmetric multipartite quantum resource.

{\bf Example 3.5.} With respect to any given product basis,  the coherence of a multipartite system  can be regarded as a completely symmetric MQC, and it is a completely symmetric multipartite quantum resource.

We consider an $n$-partite  system $H_{1}\otimes H_{2}\otimes\cdots\otimes H_{n}$, where for any $s=1,2,\cdots,n$, $H_{s}$ takes a set of basis $\{|i_{s}\rangle\}$. If $\rho$ can be represented as $\rho=\mathop\sum\limits_{i_{1},i_{2},\cdot\cdot\cdot, i_{n}}p_{i_{1}i_{2}\cdot\cdot\cdot i_{n}}|i_{1}i_{2}\cdot\cdot\cdot i_{n}\rangle\langle i_{1}i_{2}\cdot\cdot\cdot i_{n}|$, we refer to $\rho$ as an incoherent state. For  multipartite  coherence,   the free states are all incoherent states. The free operations are all local incoherent channels, corresponding to quantum operations of the form $\Phi(\rho)=\mathop\sum\limits_{j_{1},\cdot\cdot\cdot, j_{n}}(K_{j_{1}}\otimes K_{j_{2}}\otimes\cdot\cdot\cdot\otimes K_{j_{n}})\rho (K_{j_{1}}^{\dag}\otimes K_{j_{2}}^{\dag}\otimes\cdot\cdot\cdot\otimes K_{j_{n}}^{\dag})$, where $K_{j_{i}}$ is the Kraus operator corresponding to the incoherent quantum channel for any $i=1,2,\cdots,n$, with specific forms as described in \cite{AGB}.
So, as a multipartite quantum resource, the multipartite coherence should have a true measure satisfying the conditions (MQCM1)-(MQCM5). Intuitively, the hierarchy relation between sub-repartitions should be $\preccurlyeq$.

In \cite{TRMG}, the measure of  multipartite  coherence was proposed:
\begin{equation} C_{l_{1}}^{(n)}(\rho)=\sum_{(i_{1},\cdot\cdot\cdot, i_{n})\neq (i_{1}',\cdot\cdot\cdot, i_{n}')}|p_{i_{1}i_{2}\cdot\cdot\cdot i_{n}}^{i_{1}'i_{2}'\cdot\cdot\cdot i_{n}'}| \tag{3.10}\end{equation}
for any $\rho=\mathop\sum\limits_{i_{1},i_{2},\cdot\cdot\cdot, i_{n},i_{1}',i_{2}',\cdot\cdot\cdot, i_{n}'}p_{i_{1}i_{2}\cdot\cdot\cdot i_{n}}^{i_{1}'i_{2}'\cdot\cdot\cdot i_{n}'}|i_{1}i_{2}\cdot\cdot\cdot i_{n}\rangle\langle i_{1}'i_{2}'\cdot\cdot\cdot i_{n}'|\in {\mathcal S}(H_{1}\otimes H_{2}\otimes\cdots\otimes H_{n})$.
Furthermore, as shown in \cite{TRMG}, $C_{l_{1}}^{(n)}$ satisfies conditions (MQCM1) and (MQCM2). And, it is easy to see that $C^{(n)}_{l_{1}}$ satisfies conditions (MQCM3) and (MQCM5).

Note that for condition (MQCM4), the hierarchy relation $\preccurlyeq$ is natural for such coherence regarded as a MQC. So we need to check $C_{l_1}^{(n)}$ satisfies the condition (CS-MQCM4) mentioned in Example 3.4.  As the basis for each subsystem $H_{i}$ is fixed, for any sub-repartition $P=P_1|P_2|\cdots|P_m$, the part system
 $H_P=H_{P_1}\otimes H_{P_2}\otimes \cdots\otimes H_{P_m}$ has the product basis which is a part of that of the whole $n$-partite  system $H_{1}\otimes H_{2}\otimes\cdots\otimes H_{n}$. Therefore, if $Q\preccurlyeq^a P$,  as stated in \cite{ZTX},  the coherence of each subsystem is less than or equal to the coherence of the entire system, that is, we have $C_{l_1}^{(r)}(\rho_Q,Q)\leq C_{l_1}^{(m)}(\rho_P,P)$ and  (CS-MQCM4a)  is satisfied. For condition (CS-MQCM4b), assume $Q\preccurlyeq^b P$. Then, for any $\rho_{1,2,\cdots,n}\in{\mathcal S}(H_1\otimes H_2\otimes\cdots\otimes H_n)$, regarding $\rho_{P}\in {\mathcal S}(H_{P_{1}}\otimes H_{P_{2}}\otimes\cdots\otimes H_{P_{m}})$  as $m$-partite state and  $\rho_{Q}=\rho_{P}\in {\mathcal S}(H_{Q_1}\otimes H_{Q_2}\otimes\cdots\otimes H_{Q_r})$ as $r$-partite state, it is clear that $C_{l_{1}}^{(r)}(\rho_{Q},Q)=C_{l_{1}}^{(m)}(\rho_{P},P)$ and hence the condition (CS-MQCM4b) is satisfied. \if false where $\rho_{Q}={\rm Tr}_{Q^c}(\rho_{1,2,\cdots,n})$ and $\rho_{P}={\rm Tr}_{P^c}(\rho_{1,2,\cdots,n})$ are the reduced states of $\rho_{1,2,\cdots,n}$ to the subsystems $Q$ and $P$, respectively.\fi The condition (CS-MQCM4c) is directly derived from conditions (CS-MQCM4a) and (CS-MQCM4b).

Therefore, $C_{l_{1}}^{(n)}$ satisfies conditions (MQCM1)-(MQCM5), so it can  be regarded as a true MQC measure of the multipartite coherence when regarded  as a  MQC. Consequently, multipartite  coherence is a completely symmetric multipartite quantum resource.

{\bf Example 3.6.} With respect to any given product basis, the quantum imaginarity  of a multipartite system can be regarded as  a completely symmetric MQC, and it is a completely symmetric multipartite quantum resource.

We consider an $n$-partite  system $H_{1}\otimes H_{2}\otimes\cdots\otimes H_{n}$, where for any $s=1,2,\cdots,n$, $H_{s}$ is equipped with an arbitrarily given basis $\{|i_{s}\rangle\}$.  For  multipartite  imaginarity, the free states are referred to as real states, corresponding to the set of quantum states with a real density matrix:
$$\mathcal{R}=\{\rho:\langle i_{1}i_{2}\cdot\cdot\cdot i_{n}|\rho|i_{1}'i_{2}'\cdot\cdot\cdot i_{n}'\rangle\in\mathbb{R}\}.$$
The set $\mathcal{R}$  is a closed convex subset of  $\mathcal{S}(H_{1}\otimes H_{2}\otimes\cdots\otimes H_{n})$.

 In this case, free operations are those local real channels, corresponding to quantum channels $\Phi(\rho)=\mathop\sum\limits_{j_{1},\cdot\cdot\cdot, j_{n}}(K_{j_{1}}\otimes K_{j_{2}}\otimes\cdot\cdot\cdot\otimes K_{j_{n}})\rho (K_{j_{1}}^{\dag}\otimes K_{j_{2}}^{\dag}\otimes\cdot\cdot\cdot\otimes K_{j_{n}}^{\dag})$ with real Kraus operators $\langle i_{s}|K_{j_{s}}|i_{s}'\rangle\in\mathbb{R}$ for any $s=1,2,\cdot\cdot\cdot, n$ \cite{AG}.

 In \cite{AG,DVS}, the robustness of    imaginarity was proposed as
\begin{equation}\mathfrak{I}_{R}^{(n)}(\rho)=\min_{\tau}\{s\geq0:\frac{\rho+s\tau}{1+s}\in\mathcal{R}\}=\frac{1}{2}\|\rho-\rho^{\rm T}\|_{1},\tag{3.11}\end{equation}
where the minimum is taken over all quantum states $\tau$ and all $s\geq0$, ${\rm T}$ denotes transposition and $\|M\|_{1}={\rm Tr}\sqrt{M^{\dag}M}$ is the trace norm. Furthermore, as shown in \cite{AG}, $\mathfrak{I}_{R}^{(n)}$ satisfies conditions (MQCM1) and (MQCM2). And, it is easy to see that $\mathfrak{I}_{R}^{(n)}$  satisfies conditions (MQCM3) and (MQCM5). For condition (MQCM4), we will show that $\mathfrak{I}_{R}^{(n)}$ satisfies the condition (CS-MQCM4). To do this, we need only prove the following three propositions 3.5-3.7.

{\bf Proposition 3.5.} $\mathfrak{I}_{R}^{(n)}$ satisfies the property (CS-MQCM4a).

{\bf Proof.} Without loss of generality, it suffices to show that, for any $\rho_{1,2,\cdots,n}\in{\mathcal S}(H_1\otimes H_2\otimes\cdots\otimes H_n)$ and a subset $\{1,2,\cdots, r\}\subseteq\{1,2,\cdots,n\}$, we have
$$\mathfrak{I}_{R}^{(r)}(\rho_{1,2,\cdots, r})\leq \mathfrak{I}_{R}^{(n)}(\rho_{1,2,\cdots,n}),$$
 where $\rho_{1,2,\cdots,r}={\rm Tr}_{\{r+1,r+2,\cdots,n\}}(\rho_{1,2,\cdots,n})$ is the reduced state of $\rho_{1,2,\cdots,n}$ to the subsystem $\{1,2,\cdots,r\}$.

By \cite{RAE}, for any $\rho_{AB}\in\mathcal{S}(H_{A}\otimes H_{B})$ and $\rho_{A}={\rm Tr}_{B}(\rho_{AB})\in\mathcal{S}(H_{A})$, we have $\|\rho_{A}\|_{1}\leq\|\rho_{AB}\|_{1}$. So
$$\begin{array}{rl}
\mathfrak{I}_{R}^{(r)}(\rho_{1,2,\cdots, r})=&\frac{1}{2}\|\rho_{1,2,\cdots, r}-\rho_{1,2,\cdots, r}^{\rm T}\|_{1}\nonumber\\=& \frac{1}{2}\|{\rm Tr}_{\{r+1,r+2,\cdots,n\}}(\rho_{1,2,\cdots,n}-\rho_{1,2,\cdots,n}^{\rm T})\|_{1}\nonumber\\ \leq&\frac{1}{2}\|\rho_{1,2,\cdots,n}-\rho_{1,2,\cdots,n}^{\rm T}\|_{1}\nonumber\\ =& \mathfrak{I}_{R}^{(n)}(\rho_{1,2,\cdots,n})
\nonumber
\end{array}$$
as desired. \hfill$\Box$

{\bf Proposition 3.6.} $\mathfrak{I}_{R}^{(n)}$ satisfies the property (CS-MQCM4b).

{\bf Proof.}  For any $r$-partition $Q=Q_1|Q_2|\cdots |Q_r$ of $\{1,2,\cdots,n\}$ and any $\rho=\rho_{1,2,\cdots,n}\in {\mathcal S}(H_{1}\otimes H_{2}\otimes\cdots\otimes H_{n})$,  regarding $\rho$ as a $r$-partite state $\rho\in {\mathcal S}(H_{Q_1}\otimes H_{Q_2}\otimes\cdots\otimes H_{Q_r})$, we always have
$$\mathfrak{I}_{R}^{(k)}(\rho,Q)= \mathfrak{I}_{R}^{(n)}(\rho),$$
which completes the proof. \hfill$\Box$

{\bf Proposition 3.7.} $\mathfrak{I}_{R}^{(n)}$ satisfies the property (CS-MQCM4c).

{\bf Proof.}   Without loss of generality, let $P=P_1|P_2|\cdots |P_m$ be an $m$-partition of $\{1,2,\cdots,n\}$ and $Q=Q_1|Q_2|\cdots |Q_m$ such that $Q_j\subseteq P_j$.  Then, for any $\rho=\rho_{1,2,\cdots,n}\in{\mathcal S}(H_1\otimes H_2\otimes\cdots\otimes H_n)$,    $\rho\in{\mathcal S}(H_{P_1}\otimes H_{P_2}\otimes\cdots\otimes H_{P_m})$ and $\rho_Q={\rm Tr}_{Q^c}(\rho)\in{\mathcal S}(H_{Q_1}\otimes H_{Q_2}\otimes\cdots\otimes H_{Q_m})$ can be regarded  as   $m$-partite states with respect to $P$ and $Q$, respectively.

For any $\rho_{AB}\in\mathcal{S}(H_{A}\otimes H_{B})$, as $\rho_{A}={\rm Tr}_{B}(\rho_{AB})\in\mathcal{S}(H_{A})$ implies   $\|\rho_{A}\|_{1}\leq\|\rho_{AB}\|_{1}$ by \cite{RAE}, we have
$$\begin{array}{rl}
\mathfrak{I}_{R}^{(k)}(\rho_{Q},Q)=&\frac{1}{2}\|\rho_{Q}-\rho_{Q}^{\rm T}\|_{1}\nonumber\\=& \frac{1}{2}\|{\rm Tr}_{Q^c}(\rho-\rho^{\rm T})\|_{1}\nonumber\\ \leq&\frac{1}{2}\|\rho-\rho^{\rm T}\|_{1}\nonumber\\ =& \mathfrak{I}_{R}^{(k)}(\rho,P)
\nonumber
\end{array}$$
as desired. \hfill$\Box$

Therefore, the robustness $\mathfrak{I}_{R}^{(n)}$ of   multipartite  imaginarity  satisfies conditions (MQCM1)-(MQCM5), making it a true MQC measure of multipartite imaginarity, which ensures that the multipartite  imaginarity is also a MQC and a completely symmetric multipartite quantum resource.

\subsection{Gaussian multipartite quantum resources}

Note that in many quantum protocols, the systems considered are continuous-variable (CV) systems. Therefore, studying resource theory of  multipartite quantum correlations in CV systems is also very important and interesting. Recall that, an $n$-mode CV system is described by a Hilbert space $H=H_{1}\otimes H_{2}\otimes\cdots\otimes H_{n}$, where $H_{k}$ is infinite-dimensional for each $k$th mode, $k=1,2,\cdots,n$. We denote the one-mode Fock basis of $H_{k}$ by
$\{|j_{k}\rangle\}_{j_{k}=0}^{\infty}$, $k=1,2,\cdots,n$. Then the $n$-mode Fock basis of $H$ is $\{|j_{1}\rangle\otimes|j_{2}\rangle\otimes\cdots\otimes|j_{n}\rangle\}_{j_{1},j_{2},\cdots,j_{n} =0}^{\infty}$.

For an arbitrary state $\rho$
in an $n$-mode CV system, $\rho$ can be characterized by its characteristic function $\chi_{\rho}(\bm{\alpha})={\rm Tr}[\rho\mathcal{W}(\bm{\alpha})]$. $\rho$
is called a Gaussian state  if its characteristic function
 has the form
$$\chi_{\rho}(\bm{\alpha})=\exp[-\frac{1}{4}\bm{\alpha}^{\rm T}\Gamma \bm{\alpha}+i{\mathbf
d}^{\rm T}\bm{\alpha}],
$$
where $\bm{\alpha}=(x_{1},y_{1},\cdots,x_{n},y_{n})^{\rm T}\in\mathbb{R}^{2n}$,  $\mathcal{W}(\bm{\alpha})=\exp[iR^{\rm T}\bm{\alpha}]$ is the Weyl operator,
 $$\begin{array}{rl} {\mathbf d}=({\rm Tr}(\rho
\hat{R}_1), {\rm Tr}(\rho \hat{R}_2), \cdots, {\rm Tr}(\rho \hat{R}_{2n}))^{\rm
T}\in{\mathbb R}^{2n}\end{array}$$ is  the mean vector of $\rho$ and
$\Gamma=(\gamma_{kl})\in \mathcal{M}_{2n}(\mathbb R)$ is the covariance matrix of $\rho$ defined by $$\gamma_{kl}={\rm Tr}[\rho
(\Delta\hat{R}_k\Delta\hat{R}_l+\Delta\hat{R}_l\Delta\hat{R}_k)],$$
in which, $\Delta\hat{R}_k=\hat{R}_k-\langle\hat{R}_k\rangle$,
$\langle\hat{R}_k\rangle={\rm Tr}[\rho\hat{R}_k]$,
$R=(\hat{R}_1,\hat{R}_2,\cdots,\hat{R}_{2n})=(\hat{Q}_1,\hat{P}_1,\cdots,\hat{Q}_n,\hat{P}_n)$.
$\hat{Q_k}=(\hat{a}_k+\hat{a}_k^\dag)/\sqrt{2}$,
$\hat{P_k}=-i(\hat{a}_k-\hat{a}_k^\dag)/\sqrt{2}$ ($k=1,2,\cdots,n$)
are  respectively the position and momentum operators in the $k$th mode,
$\hat{a}_k^\dag$ and $\hat{a}_k$ are respectively the creation and annihilation
operators in the $k$th mode \cite{SP,GJI}. Here, as usual, $\mathcal{M}_d(\mathbb R)$
stands for the algebra of all $d\times d$ matrices over the real
field $\mathbb R$. Therefore, every Gaussian state $\rho$ is determined by its covariance matrix $\Gamma$ and mean vector ${\mathbf d}$, and thus, one can write $\rho=\rho(\Gamma, {\mathbf d})$.

Below, we present some completely symmetric multipartite Gaussian quantum resources.

{\bf Example 3.7.} The multipartite multi-mode Gaussian non-product correlation  is a completely symmetric multipartite Gaussian quantum resource.

  In an $(m_{1}+m_{2}+\cdots+m_{n})$-mode $n$-partite CV system with state space $H_{1}\otimes H_{2}\otimes\cdots\otimes H_{n}$, the set of Gaussian product states  forms a closed set. Consequently, for Gaussian non-product correlations, the free states are identified as Gaussian product states. In other words, the set of free states consists of Gaussian states $\rho$ that can be expressed as $\rho=\rho_{1}\otimes\rho_{2}\otimes\cdots\otimes\rho_{n}$, where each $\rho_{i}$ is a $m_i$-mode Gaussian state in the system $H_{i}$ for any $i=1,2,\cdots,n$. The free operations correspond to local Gaussian channels of the form $\Phi=\Phi_{1}\otimes\Phi_{2}\otimes\cdot\cdot\cdot\otimes\Phi_{n}$. So, as a multipartite Gaussian quantum resource, it should have a true measure of  $n$-partite multi-mode Gaussian non-product satisfying the conditions (MQCM1)-(MQCM5).

 In \cite{HLQ22}, a measure of $n$-partite multi-mode Gaussian non-product for CV systems was proposed:
\begin{equation}\mathcal{M}^{(n)}(\rho_{1,2,\cdots,n})=1-\frac{{\rm det}(\Gamma_{\rho_{1,2,\cdots,n}})}{\Pi_{j=1}^{n}{\rm det}(\Gamma_{\rho_{j}})} \tag{3.12}\end{equation}
for any $(m_{1}+m_{2}+\cdots+m_{n})$-mode $n$-partite Gaussian state $\rho_{1,2,\cdots,n}\in\mathcal{S}(H_{1}\otimes H_{2}\otimes\cdots\otimes H_{n})$, where $\Gamma_{\rho_{1,2,\cdots,n}}$ and $\Gamma_{\rho_{j}}$ are respectively the covariance matrices of $\rho_{1,2,\cdots,n}$ and $\rho_{j}={\rm Tr}_{j^{c}}(\rho_{1,2,\cdots,n})$, with $j^{c}=\{1,2,\cdots,n\}\backslash \{j\}$ representing the complement of the index $j$ in the set $\{1,2,\cdots,n\}$. Furthermore, as shown in \cite{HLQ22}, $\mathcal{M}^{(n)}$ satisfies conditions (MQCM1)-(MQCM5) with $\preccurlyeq$ as the hierarchy relation between sub-repartitions, making it a true MQC measure
of multipartite multi-mode Gaussian non-product. Therefore, the multipartite multi-mode Gaussian non-product correlation is a type of completely symmetric multipartite Gaussian quantum resource.

{\bf Example 3.8.} With respect to the product Fock basis, the multipartite multi-mode Gaussian imaginarity  is a completely symmetric multipartite Gaussian quantum resource.

For an $n$-mode CV  system $H=H_{1}\otimes H_{2}\otimes\cdots\otimes H_{n}$, if a Gaussian state $\rho$ satisfies
$$\langle j_{1}|\langle j_{2}|\cdots\langle j_{n}|\rho|i_{1}\rangle|i_{2}\rangle\cdots|i_{n}\rangle\in\mathbb{R}$$
for all Fock basis vectors $|j_{1}\rangle,|j_{2}\rangle,\cdots,|j_{n}\rangle$ and $ |i_{1}\rangle,|i_{2}\rangle,\cdots,|i_{n}\rangle$, then $\rho$ is called a real Gaussian state \cite{XJW}. Additionally, for any $n$-mode Gaussian state $\rho(\Gamma_{\rho}, {\mathbf d}_{\rho})$, an easily computable measure of  multipartite  Gaussian imaginarity was proposed in \cite{ZHQ}:
\begin{equation}\mathcal{I}^{G_{n}}(\rho)=1-\frac{{\rm det}(\Gamma_{\rho})}{{\rm det}(Q_{n}P_{n}\Gamma_{\rho}P_{n}^{\rm T}Q_{n}^{\rm T}){\rm det}(Q_{n}'P_{n}\Gamma_{\rho}P_{n}^{\rm T}Q_{n}'^{\rm T})}+h(\|Q_{n}'P_{n}{\mathbf d}_{\rho}\|_{1}),\tag{3.13}\end{equation}
where $P_{n}=(p_{ij})_{2n\times2n}\in\mathcal{M}_{2n}(\mathbb R)$ is a permutation matrix satisfying $p_{i,2i-1}=p_{n+i,2i}=1$ for any $i=1,2,\cdots,n$ and other elements 0, $Q_{n}=(I_{n},0)_{n\times 2n}$ and $Q_{n}'=(0,I_{n})_{n\times 2n}$. Additionally, $h:[0,+\infty)\rightarrow\{0,1\}$ is a function with $h(z)=0$ if $z=0$ and $h(z)=1$ if $z\neq0$.

In an $(m_{1}+m_{2}+\cdots+m_{n})$-mode $n$-partite CV system with state space $H=H_{1}\otimes H_{2}\otimes\cdots\otimes H_{n}$,  the set of real Gaussian states forms a closed set. Consequently, for Gaussian imaginarity, the free states are identified as real Gaussian  states.

The free operations correspond to local real Gaussian channels of the form $\Phi=\Phi_{1}\otimes\Phi_{2}\otimes\cdot\cdot\cdot\otimes\Phi_{n}$, where each $\Phi_{k}$ is either completely real or covariant real, the specific forms of which are shown in \cite{XJW,ZHQ}.

Additionally, for any $(m_{1}+m_{2}+\cdots+m_{n})$-mode $n$-partite Gaussian state $\rho$, $\mathcal{I}^{G_{m}}_{n}(\rho)=\mathcal{I}^{G_{m}}(\rho)$ is a Gaussian imaginarity measure, where $m=m_{1}+m_{2}+\ldots+m_{n}$ is the mode of $H$. Furthermore, as shown in \cite{ZHQ}, $\mathcal{I}^{G_{m}}_{n}$ satisfies conditions (MQCM1)-(MQCM5) with $\preccurlyeq$ as the hierarchy relation between sub-repartitions, making it a true measure
of multipartite multi-mode Gaussian imaginarity.

Therefore, the multipartite multi-mode Gaussian imaginarity can be regarded as a MQC and forms also a  completely symmetric multipartite Gaussian quantum resource.

\subsection{Other symmetric MQCs}

A MQC may be symmetric about subsystems but not completely symmetric. \if false In these situations, the hierarchy condition for a true MQC measure may be part of (CS-MQCM4a)-(CS-MQCM4c), according to what MQC is.\fi

{\bf Example 3.9.} With respect to the product Fock basis, the multipartite single-mode Gaussian coherence is a  symmetric multipartite Gaussian quantum resource.

Consider an $n$-partite CV system $H=H_{1}\otimes H_{2}\otimes\cdots\otimes H_{n}$, where each subsystem $H_{i}$ is a single-mode system. Under this constraint, the hierarchy condition for a true measure of multipartite single-mode Gaussian coherence is limited to type (CS-MQCM4a).

For $n$-partite single-mode Gaussian coherence of CV systems, the free states are  Gaussian incoherent states, and the corresponding set is
$$\mathcal{GI}=\{\rho(\Gamma_{\rho},{\mathbf d}_\rho):\Gamma_{\rho}=\oplus_{k=1}^n
\lambda_k I_2,{\mathbf d}_\rho=0\}.$$
The set $\mathcal{GI}$ is a closed set.

The free operations are local Gaussian quantum incoherent channels, that is, $\Phi=\Phi_{1}\otimes\Phi_{2}\otimes\cdot\cdot\cdot\otimes\Phi_{n}$, where each $\Phi_{i}$ is single-mode  Gaussian incoherent channel, the specific forms of which are shown in \cite{HHQ24}.

In \cite{HHQ24}, a measure of  multipartite single-mode Gaussian coherence for CV systems was proposed:
\begin{equation} C_\nu^{G_n} (\rho)= 1-\frac{2^n{\rm
det}(\Gamma_\rho)}{\prod_{k=1}^n {\rm Tr}(V_{kk}^2)}+\sum_{k=1}^n h(\| w_k\|_2) \tag{3.14}\end{equation}
for any Gaussian state $\rho=\rho_{1,2,\cdots,n}\in{\mathcal S}(H_{1}\otimes H_{2}\otimes\cdots\otimes H_{n})$  with covariance matrix $\Gamma_\rho=(v_{ij})_{2n \times 2n}\in{\mathcal M}_{2n}(\mathbb R)$ and mean vector ${\mathbf d}_\rho=(d_1, d_2, \cdots, d_{2n})^{\rm T}\in{\mathbb R}^{2n}$, where $V_{ij}=\left(\begin{array}{cc} v_{2i-1,2j-1} & v_{2i-1,2j} \\ v_{2i,2j-1} & v_{2i,2j} \end{array}\right)$ and $w_i=(d_{2i-1}, d_{2i})^{\rm T}$, $i,j=1,2,\cdots,n$. Furthermore, as shown in \cite{HHQ24}, $C_\nu^{G_n}$ satisfies conditions (MQCM1)-(MQCM3) and (MQCM5) as well as (CS-MQCM4a). It follows that  $C_\nu^{G_n}$ is a true MQC measure of the multipartite single-mode  Gaussian coherence. Therefore, multipartite single-mode Gaussian coherence of CV systems is a symmetric multipartite Gaussian quantum resource.

With respect to the product Fock basis, it remains unclear whether multipartite multi-mode Gaussian coherence qualifies  as a multipartite Gaussian quantum resource. The main challenge lies in the fact that no true measure of multipartite multi-mode Gaussian coherence satisfying conditions (MQCM1)-(MQCM5) has been discovered yet.

\section{Asymmetric multipartite quantum resources}

In addition to the symmetric MQCs discussed in the previous section, there are also  asymmetric MQCs that hold significant importance in quantum communication and quantum computing, such as multipartite quantum steering and
multipartite quantum discord.

A MQC is said to be asymmetric if it is defined for any $n$-partite system $H=H_1\otimes H_2\otimes \cdots\otimes H_n$ of any dimension and it is asymmetric with respect to changing orders of the subsystems. This means that, there exists a  $\rho\in {\mathcal S}(H_1\otimes H_2\otimes \cdots\otimes H_n)$ and a permutation $\pi$ of $(1,2,\cdots,n)$, so that $\rho$  is MQC, but $\rho^{\pi}\in{\mathcal S}(H_{{\pi(1)}}\otimes H_{{\pi(2)}}\otimes \cdots\otimes H_{{\pi(n)}})$ is non-MQC, where $\rho^{\pi}$ is obtained from $\rho$ by changing the orders of subsystems $H_j$ according to $\pi$.

In this situation, a true asymmetric MQC measure $\mathcal{C}^{(n)}$ should satisfy conditions (MQCM1)-(MQCM3), as well as the hierarchy condition determined by the multipartite quantum correlation itself.

The main purpose of this section is to show the following

{\bf Theorem 4.1.} The  multipartite steering  is an  asymmetric multipartite quantum resource.

Let us review the definition of multipartite quantum steering.
A measurement assemblage  ${\mathcal {MA}} =
\{M_{a|x}\}_{a,x}$ is a collection of positive operators $M_{a|x}
\geq0$ satisfying $\sum_a M_{a|x} = I$ for each $x$. Such a
collection represents one positive-operator-valued measurement (POVM) for each $x$.  For multipartite quantum systems, Cavalcanti
et al. \cite{EQM} proposed the definition of multipartite steering correlation for $n$ spatially separated parties among which $t$ are
untrusted and $n-t$ are trusted. Each untrusted party $A_{i}$  performs
a set of measurements
$\{M^{A_{i}}_{a_{i}|x_{i}}\}_{a_{i},x_{i}}$, $i=1,2,\cdots,t$, and then the collection of sub-normalized ``conditional states" of $A_{t+1}\otimes A_{t+2}\otimes\cdot\cdot\cdot\otimes A_{n}$ forms an  assemblage
$\{\sigma_{a_{1},\cdots,a_{t}|x_{1},\cdots,x_{t}}\}_{a_{1},\cdots,a_{t},x_{1},\cdots,x_{t}}$ with
\begin{equation}\label{eq4.1}\sigma_{a_{1},\cdots,a_{t}|x_{1},\cdots,x_{t}}={\rm Tr}_{\{1,\cdots,t\}}\{[M^{A_{1}}_{a_{1}|x_{1}}\otimes\cdots\otimes M^{A_{t}}_{a_{t}|x_{t}}\otimes I_{t+1}\otimes\cdots\otimes I_{n}]\rho_{1,2,\cdots,n}\}.\tag{4.1}\end{equation}
It clear that $\mathop{\sum}\limits_{a_{1},a_{2},\cdots,a_{t}}\sigma_{a_{1},\cdots,a_{t}|x_{1},\cdots,x_{t}}=\rho_{t+1,t+2,\cdots,n}$ for any $x_{1},x_{2},\cdots,x_{t}$, where $\rho_{t+1,t+2,\cdots,n}={\rm Tr}_{\{1,\cdots,t\}}(\rho_{1,2,\cdots,n})$.

The state $\rho_{1,2,\cdots,n}$ is said to be  unsteerable from  $(A_{1}A_{2}\cdot\cdot\cdot A_{t})$ to $(A_{t+1}A_{t+2}\cdot\cdot\cdot A_{n})$  if every assemblage $\{\sigma_{a_{1},\cdots,a_{t}|x_{1},\cdots,x_{t}}\}_{a_{1},\cdots,a_{t},x_{1},\cdots,x_{t}}$ on  $A_{t+1}\otimes A_{t+2}\otimes\cdot\cdot\cdot\otimes A_{n}$ can be explained by a  local hidden state
(LHS) model  as follows:
\begin{equation}\label{eq4.2}\sigma_{a_{1},\cdots,a_{t}|x_{1},\cdots,x_{t}}=\sum_{\lambda}P(\lambda)\Pi_{i=1}^{t}P(a_{i}|x_{i},\lambda)
\sigma_{\lambda}^{t+1}\otimes\sigma_{\lambda}^{t+2}\otimes\cdot\cdot\cdot\otimes\sigma_{\lambda}^{n},\tag{4.2}\end{equation}
where $P(\lambda)$ is  the probability distribution with respect to $\lambda$, the quantum states $\{\sigma_\lambda^j\}_\lambda$ of $A_{j}$ $(j=t+1,t+2,\cdots,n)$ are predetermined. Otherwise, $\rho$ is steerable from  $(A_{1}A_{2}\cdot\cdot\cdot A_{t})$ to $(A_{t+1}A_{t+2}\cdot\cdot\cdot A_{n})$. \if false For
a given $\lambda$, their state is $\sigma_{\lambda}^{t+1}\otimes\sigma_{\lambda}^{t+2}\otimes\cdot\cdot\cdot\otimes\sigma_{\lambda}^{n}$.\fi By Eq.(\ref{eq4.2}), if $\rho_{1,2,\cdots,n}$ is unsteerable from  $(A_{1}A_{2}\cdot\cdot\cdot A_{t})$ to $(A_{t+1}A_{t+2}\cdot\cdot\cdot A_{n})$ and $n-t\geq2$, then
$$\rho_{t+1,t+2,\cdots,n}=\sum_{a_{1},a_{2},\cdots,a_{t}}\sigma_{a_{1},\cdots,a_{t}|x_{1},\cdots,x_{t}}=\sum_{\lambda}P(\lambda)\sigma_{\lambda}^{t+1}\otimes\sigma_{\lambda}^{t+2}\otimes\cdot\cdot\cdot\otimes\sigma_{\lambda}^{n}$$
is fully separable, where $\rho_{t+1,t+2,\cdots,n}={\rm Tr}_{\{1,\cdots,t\}}(\rho_{1,2,\cdots,n})$.

 Denote by $\mathcal{US}_{(1\cdot\cdot\cdot t)\rightarrow (t+1\cdot\cdot\cdot n)}$ the set of all unsteerable states from $(A_{1}A_{2}\cdot\cdot\cdot A_{t})$ to $(A_{t+1}A_{t+2}\cdot\cdot\cdot A_{n})$, which is clearly a closed convex subset of  $\mathcal{S}(H_{1}\otimes H_{2}\otimes\cdots\otimes H_{n})$ and then regarded as the set of all free states.

The free operations are defined by quantum channels
\begin{equation}\label{eq4.3}
 \Phi(\rho) =\sum_j K_j\rho K_j^\dag
\tag{4.3}\end{equation}
with the Kraus operators of the form $K_j=I_{1}\otimes\cdots\otimes I_{t}\otimes K_{j_{t+1}}\otimes K_{j_{t+2}}\otimes\cdot\cdot\cdot\otimes K_{j_{n}}$ and $\sum_j K_j^\dag K_j=I$.


So, if this multipartite steering correlation is  a multipartite quantum resource, one should have a true multipartite steering measure satisfying the conditions (MQCM1)-(MQCM4), but here we write them as (MStM1)-(MStM4).

Assume that  $S^{(1\cdot\cdot\cdot t)\rightarrow (t+1\cdot\cdot\cdot n)}$ is a true
 steering measure from $(A_{1}A_{2}\cdot\cdot\cdot A_{t})$ to $(A_{t+1}A_{t+2}\cdot\cdot\cdot A_{n})$; then, for any $\rho=\rho_{1,2,\ldots,n}\in\mathcal S(H_1\otimes H_2\otimes\cdots\otimes H_n)$, it should satisfy the following axioms.

 {\bf (MStM1)} (Faithfulness) $S^{(1\cdot\cdot\cdot t)\rightarrow (t+1\cdot\cdot\cdot n)}(\rho)\geq 0$ and  $S^{(1\cdot\cdot\cdot t)\rightarrow (t+1\cdot\cdot\cdot n)}(\rho)= 0$ if and only if $\rho$ is unsteerable from
 $(A_1A_2\cdots A_t)$ to $(A_{t+1}A_{t+2}\ldots A_n)$.

{\bf (MStM2)} (Monotonicity under free operations) For any  channel  $\Phi$ of the form in Eq.(\ref{eq4.3}), we have $$S^{(1\cdot\cdot\cdot t)\rightarrow (t+1\cdot\cdot\cdot n)}(\Phi(\rho))\leq S^{(1\cdot\cdot\cdot t)\rightarrow (t+1\cdot\cdot\cdot n)}(\rho).$$

However, for multipartite steering, both  the unification condition and the hierarchy condition  are somewhat complicated. Naturally, the unification condition requires that, for any subsets $\emptyset \neq X \subseteq \{1,2,\cdots,t\}$ and $\emptyset \neq Y \subseteq \{t+1,t+2,\cdots,n\}$, the steering measure $S^{X\rightarrow Y}$ from  $X$ to $Y$  should be defined.  So the unification for a steering measure $S^{(1\cdot\cdot\cdot t)\rightarrow (t+1\cdot\cdot\cdot n)}$ reads as follows.

{\bf (MStM3)} (Unification condition)
 $\{S^{X\rightarrow Y} : \emptyset \neq X \subseteq \{1,2,\cdots,t\} \text{ and } \emptyset \neq Y \subseteq \{t+1,t+2,\cdots,n\}\}$ are defined in the same way such that $S^{X\rightarrow Y}$ can be used to measure the steering in $H_{XY}=\bigotimes _{i\in X} H_i\otimes \bigotimes_{j\in Y} H_j$ from $\{A_i, i\in X\}$ to $\{A_j: j\in Y\}$ without causing confusions.

Let $\mathcal{SP}_{(1\rightarrow t)}$ and $\mathcal{SP}_{(t+1\rightarrow n)}$ denote the sets of all sub-repartitions of $\{1,2,\cdots, t\}$ and $\{t+1,t+2,\cdots, n\}$, respectively. The hierarchy condition for multipartite steering measure ensures that, for any sub-repartitions $P=P_1|P_2|\cdots|P_m, P'=P_1'|P_2'|\cdots |P_r'\in\mathcal{SP}_{(1\rightarrow t)}$ and  $Q=Q_1|Q_2|\cdots|Q_s, Q'=Q_1'|Q_2'|\cdots |Q_l'\in\mathcal{SP}_{(t+1\rightarrow n)}$, if  the hierarchy of
$(P,Q)$ is lower than that of $(P',Q')$ with respect to steering, denoted by $(P,Q)\preccurlyeq_{s} (P',Q')$, then for any quantum state $\rho\in {\mathcal S}(H_{1}\otimes H_{2}\otimes\cdots\otimes H_{n})$, the  steering of its reduced state $\rho_N$ with respect to the part system $H_N=H_{P_1}\otimes\cdots\otimes H_{P_m}\otimes H_{Q_1}\otimes\cdots\otimes H_{Q_s}$ never exceeds the  steering of its reduced state $\rho_M$ with respect to the part system $H_M=H_{P_1'}\otimes\cdots\otimes H_{P_r'}\otimes H_{Q_1'}\otimes\cdots\otimes H_{Q_l'}$. Thus, to clarify the conception of the hierarchy condition for multipartite steering measure, we should first give the exact meaning of $(P,Q)\preccurlyeq_{s} (P',Q')$. If $(P,Q)\preccurlyeq_{s} (P',Q')$, we also say that $(P,Q)$ is coarser than $(P',Q')$ with respect to steering, for simplicity.

{\bf Definition 4.2.} For any sub-repartitions $P=P_1|P_2|\cdots|P_m, \ P'=P_1'|P_2'|\cdots |P_r'\in\mathcal{SP}_{(1\rightarrow t)}$ and  $Q=Q_1|Q_2|\cdots|Q_s,\ Q'=Q_1'|Q_2'|\cdots |Q_l'\in\mathcal{SP}_{(t+1\rightarrow n)}$, there are nine basic types of hierarchy such that $(P,Q)$ is coarser than $(P',Q')$ with respect to steering:

(s-a-a)  $P\preccurlyeq^{a}P'$ and $Q\preccurlyeq^{a}Q'$,   denoted by $(P,Q)\preccurlyeq_{s}^{(a,a)} (P',Q')$;

(s-a-b) $P\preccurlyeq^{a}P'$ and $Q\preccurlyeq^{b}Q'$,   denoted by  $(P,Q)\preccurlyeq_{s}^{(a,b)} (P',Q')$;

(s-a-c) $P\preccurlyeq^{a}P'$ and $Q\preccurlyeq^{c}Q'$, denoted  by  $(P,Q)\preccurlyeq_{s}^{(a,c)} (P',Q')$;

(s-b-a) $P'\preccurlyeq^{b}P$ and $Q\preccurlyeq^{a}Q'$,  denoted  by  $(P,Q)\preccurlyeq_{s}^{(b,a)} (P',Q')$;

(s-b-b)  $P'\preccurlyeq^{b}P$ and $Q\preccurlyeq^{b}Q'$,   denoted  by  $(P,Q)\preccurlyeq_{s}^{(b,b)} (P',Q')$;

(s-b-c) $P'\preccurlyeq^{b}P$ and $Q\preccurlyeq^{c}Q'$, denoted by  $(P,Q)\preccurlyeq_{s}^{(b,c)} (P',Q')$;

(s-c-a) $P\preccurlyeq^{c}P'$ and $Q\preccurlyeq^{a}Q'$,   denoted  by  $(P,Q)\preccurlyeq_{s}^{(c,a)} (P',Q')$;

(s-c-b) $P\preccurlyeq^{c}P'$ and $Q\preccurlyeq^{b}Q'$,  denoted  by  $(P,Q)\preccurlyeq_{s}^{(c,b)} (P',Q')$;

(s-c-c) $P\preccurlyeq^{c}P'$ and $Q\preccurlyeq^{c}Q'$,   denoted  by  $(P,Q)\preccurlyeq_{s}^{(c,c)} (P',Q')$.

Now, for any sub-repartitions $P=P_1|P_2|\cdots|P_m,\ P'=P_1'|P_2'|\cdots |P_r'\in\mathcal{SP}_{(1\rightarrow t)}$ and $Q=Q_1|Q_2|\cdots|Q_s,\ Q'=Q_1'|Q_2'|\cdots |Q_l'\in\mathcal{SP}_{(t+1\rightarrow n)}$, we say that the hierarchy of $(P,Q)$ is lower than the hierarchy of   $(P',Q')$ with respect to steering, or, in other words, $(P,Q)$ is coarser than  $(P',Q')$ with respect to steering, denoted by $(P,Q)\preccurlyeq_{s} (P',Q')$,   if   there exist sub-repartitions $R_1,R_2,\cdots, R_z \in\mathcal{SP}_{(1\rightarrow t)}$ and $G_1,G_2,\cdots, G_z \in\mathcal{SP}_{(t+1\rightarrow n)}$ such that
\begin{equation}\label{eq4.0}(P,Q)\preccurlyeq^{(x_1,y_{1})}_{s} (R_{1},G_{1})\preccurlyeq^{(x_2,y_{2})}_{s} (R_{2},G_{2}) \preccurlyeq^{(x_3,y_{3})}_{s}\cdots \preccurlyeq^{(x_z,y_{z})}_{s} (R_{z},G_{z})\preccurlyeq^{(x_{z+1},y_{z+1})}_{s} (P',Q'),\tag{4.4}\end{equation}
where $x_1,x_2,\cdots,x_z,x_{z+1}\in\{a,b,c\}$ and $y_1,y_2,\cdots,y_z,y_{z+1}\in\{a,b,c\}$.

Note that the basic relation (s-b-x) with $x\in\{a,b,c\}$ seems strange. However, this is reasonable because $P'\preccurlyeq^b P$ means that one has more choices of local measurements on $H_{P'}$ than that on $H_P$ and hence a state $\rho_{P'}=\rho_P$ is more easily to be steerable from $P'$ to $Q'$.

Thus  the hierarchy condition for a steering measure $S^{(1\cdot\cdot\cdot t)\rightarrow (t+1\cdot\cdot\cdot n)}$ can be rewritten as follows.

{\bf (MStM4)} (Hierarchy condition) $S^{(1\cdot\cdot\cdot t)\rightarrow (t+1\cdot\cdot\cdot n)}$ has  the following nine properties: for any $\rho=\rho_{1,2,\cdots,n}\in{\mathcal S}(H_1\otimes H_2\otimes\cdots\otimes H_n)$, and any sub-repartitions $P=P_1|P_2|\cdots|P_m,\ P'=P_1'|P_2'|\cdots |P_r'\in\mathcal{SP}_{(1\rightarrow t)}$ and $Q=Q_1|Q_2|\cdots|Q_s,\ Q'=Q_1'|Q_2'|\cdots |Q_l'\in\mathcal{SP}_{(t+1\rightarrow n)}$,

(MStM4aa) $(P,Q)\preccurlyeq_{s}^{(a,a)} (P',Q')$ implies $S^{P\rightarrow Q}(\rho_{N}, N)\leq S^{P'\rightarrow Q'}(\rho_{M},M)$;

(MStM4ab) $(P,Q)\preccurlyeq_{s}^{(a,b)} (P',Q')$ implies $S^{P\rightarrow Q}(\rho_{N}, N)\leq S^{P'\rightarrow Q'}(\rho_{M},M)$;

(MStM4ac) $(P,Q)\preccurlyeq_{s}^{(a,c)} (P',Q')$ implies $S^{P\rightarrow Q}(\rho_{N}, N)\leq S^{P'\rightarrow Q'}(\rho_{M},M)$;

(MStM4ba) $(P,Q)\preccurlyeq_{s}^{(b,a)} (P',Q')$ implies $S^{P\rightarrow Q}(\rho_{N}, N)\leq S^{P'\rightarrow Q'}(\rho_{M},M)$;

(MStM4bb) $(P,Q)\preccurlyeq_{s}^{(b,b)} (P',Q')$ implies $S^{P\rightarrow Q}(\rho_{N}, N)\leq S^{P'\rightarrow Q'}(\rho_{M},M)$;

(MStM4bc) $(P,Q)\preccurlyeq_{s}^{(b,c)} (P',Q')$ implies $S^{P\rightarrow Q}(\rho_{N}, N)\leq S^{P'\rightarrow Q'}(\rho_{M},M)$;

(MStM4ca) $(P,Q)\preccurlyeq_{s}^{(c,a)} (P',Q')$ implies $S^{P\rightarrow Q}(\rho_{N}, N)\leq S^{P'\rightarrow Q'}(\rho_{M},M)$;

(MStM4cb) $(P,Q)\preccurlyeq_{s}^{(c,b)} (P',Q')$ implies $S^{P\rightarrow Q}(\rho_{N}, N)\leq S^{P'\rightarrow Q'}(\rho_{M},M)$;

(MStM4cc) $(P,Q)\preccurlyeq_{s}^{(c,c)} (P',Q')$ implies $S^{P\rightarrow Q}(\rho_{N}, N)\leq S^{P'\rightarrow Q'}(\rho_{M},M)$,\\
where $M=P'|Q'$, $N=P|Q$,  $\rho_{M}={\rm Tr}_{M^c}(\rho)$ and $\rho_{N}={\rm Tr}_{N^c}(\rho)$  are the reduced states of $\rho$ to the subsystems $M$ and $N$, respectively.

Now we propose a quantification $\mathbb S^{(1\cdot\cdot\cdot t)\rightarrow (t+1\cdot\cdot\cdot n)}$ of the multipartite steering that satisfies the axioms (MStM1)-(MStM4).

Consider the multipartite steering from $(A_1A_2\cdots A_t)$ to $(A_{t+1}A_{t+2}\cdots A_n)$ for the $n$-partite system $H_1\otimes H_2\otimes\cdots\otimes H_n$. For any   $\rho\in\mathcal{S}(H_{1}\otimes H_{2}\otimes\cdot\cdot\cdot\otimes H_{n})$, define
\begin{equation}\label{eq4.4} \mathbb{S}^{(1\cdot\cdot\cdot t)\rightarrow (t+1\cdot\cdot\cdot n)}(\rho)=\min_{\sigma\in\mathcal{US}_{(1\cdot\cdot\cdot t)\rightarrow (t+1\cdot\cdot\cdot n)}}\frac{1}{2}\|\rho-\sigma\|_{1},\tag{4.5}\end{equation}
where $\|M\|_{1}={\rm Tr}\sqrt{M^{\dag}M}$ is the trace norm. It is evident that $\mathbb{S}^{(1\cdot\cdot\cdot t)\rightarrow (t+1\cdot\cdot\cdot n)}$ satisfy conditions (MStM1), (MStM2) and (MStM3). For condition (MStM4),  we will conduct a  verification by proving several propositions.

{\bf Proposition 4.3.} For any $\rho_{1,2,\cdots,n}\in{\mathcal S}(H_1\otimes H_2\otimes\cdots\otimes H_n)$ and  any sub-repartitions $P=P_1|P_2|\cdots|P_m,\ P'=P_1'|P_2'|\cdots |P_r'\in\mathcal{SP}_{(1\rightarrow t)}$ and  $Q'=Q=Q_1|Q_2|\cdots|Q_s\in\mathcal{SP}_{(t+1\rightarrow n)}$. If $P\preccurlyeq^{x} P'$ for $x\in\{a,c\}$ or $P'\preccurlyeq^{b} P$,  then we have
$$\mathbb{S}^{P\rightarrow Q}(\rho_{N}, N)\leq \mathbb{S}^{P'\rightarrow Q}(\rho_{M},M),$$
 where $M=P'|Q'=P'|Q$, $N=P|Q$,  $\rho_{M}={\rm Tr}_{M^c}(\rho_{1,2,\cdots,n})$ and $\rho_{N}={\rm Tr}_{N^c}(\rho_{1,2,\cdots,n})$  are the reduced states of $\rho_{1,2,\cdots,n}$ to the subsystems $M$ and $N$, respectively.

{\bf Proof.} To prove this proposition, we consider the  three cases.

{\bf Case (1)}.  $P\preccurlyeq^{a} P'$.

 Without loss of generality, assume $P'=\{1,2,\cdots,t\}$, $P=\{1,2,\cdots,m\}$ and $Q=Q'=\{t+1,t+2,\cdots,n\}$ where $m\leq t$. It suffices to prove the inequality
$$\mathbb{S}^{(1\cdot\cdot\cdot m)\rightarrow (t+1\cdot\cdot\cdot n)}(\rho_{1,\cdots, m, t+1,\cdots,n})\leq \mathbb{S}^{(1\cdot\cdot\cdot t)\rightarrow (t+1\cdot\cdot\cdot n)}(\rho_{1,\cdots, t, t+1,\cdots,n}),$$
where $\rho_{1,\cdots, m, t+1,\cdots,n}={\rm Tr}_{\{m+1,\cdots,t\}}(\rho_{1,2,\cdots,n})$.

For any $\sigma_{1,\cdots,t,t+1,\cdots,n}\in\mathcal{US}_{(1\cdot\cdot\cdot t)\rightarrow (t+1\cdot\cdot\cdot n)}$ and measurement assemblages $\{M^{A_{1}}_{a_{1}|x_{1}}\}_{a_{1},x_{1}}$, $\{M^{A_{2}}_{a_{2}|x_{2}}\}_{a_{2},x_{2}}$,$\cdot\cdot\cdot$, $\{M^{A_{t}}_{a_{t}|x_{t}}\}_{a_{t},x_{t}}$ of $A_{1}, A_{2},\cdot\cdot\cdot,A_{t}$, Eqs.(\ref{eq4.1}) and (\ref{eq4.2}) yield
\begin{equation}\tag{4.6}\label{eq4.5}
\begin{aligned}
&{\rm Tr}_{\{1,\cdots,t\}}\{[M^{A_{1}}_{a_{1}|x_{1}}\otimes\cdots\otimes M^{A_{t}}_{a_{t}|x_{t}}\otimes I_{t+1}\otimes\cdots\otimes I_{n}]\sigma_{1,\cdots,t,t+1,\cdots,n}\}\\ =&\sum_{\lambda}P(\lambda)\Pi_{i=1}^{t}P(a_{i}|x_{i},\lambda)\sigma_{\lambda}^{t+1}\otimes\sigma_{\lambda}^{t+2}\otimes\cdot\cdot\cdot\otimes\sigma_{\lambda}^{n}.
\nonumber
\end{aligned}
\end{equation}

Now consider $\sigma_{1,\cdots, m, t+1,\cdots,n}={\rm Tr}_{\{m+1,\cdots, t\}}(\sigma_{1,\cdots,t,t+1,\cdots,n})$. For any measurement assemblages $\{M^{A_{1}}_{a_{1}|x_{1}}\}_{a_{1},x_{1}}$, $\{M^{A_{2}}_{a_{2}|x_{2}}\}_{a_{2},x_{2}}$,$\cdot\cdot\cdot$, $\{M^{A_{m}}_{a_{m}|x_{m}}\}_{a_{m},x_{m}}$ of $A_{1},\cdot\cdot\cdot, A_{m}$,
\begin{equation}\tag{4.7}\label{eq4.6}
\begin{aligned}
&{\rm Tr}_{\{1,\cdots,m\}}\{(M^{A_{1}}_{a_{1}|x_{1}}\otimes\cdots\otimes M^{A_{m}}_{a_{m}|x_{m}}\otimes I_{t+1}\otimes\cdot\cdot\cdot\otimes I_{n})\sigma_{1,\cdots, m, t+1,\cdots,n}\}\\ =&{\rm Tr}_{\{1,\cdots,m\}}\{(M^{A_{1}}_{a_{1}|x_{1}}\otimes\cdots\otimes M^{A_{m}}_{a_{m}|x_{m}}\otimes  I_{t+1}\otimes\cdot\cdot\cdot\otimes I_{n}){\rm Tr}_{\{m+1,\cdots, t\}}(\sigma_{1,\cdots,t,t+1,\cdots,n})\}\\ =&{\rm Tr}_{\{m+1,\cdots, t\}}\{ {\rm Tr}_{\{1,\cdots,m\}}[(M^{A_{1}}_{a_{1}|x_{1}}\otimes\cdots\otimes M^{A_{m}}_{a_{m}|x_{m}}\otimes I_{m+1}\otimes\cdots\otimes  I_{n})\sigma_{1,\cdots,t,t+1,\cdots,n}]\}\\ =&\sum_{\lambda}P(\lambda)\Pi_{i=1}^{m}P(a_{i}|x_{i},\lambda)\sigma_{\lambda}^{t+1}\otimes\sigma_{\lambda}^{t+2}\otimes\cdot\cdot\cdot\otimes\sigma_{\lambda}^{n}.
\end{aligned}
\end{equation}
The final equality in inequality (\ref{eq4.6}) follows from Eq.(\ref{eq4.5}) by setting $M^{A_{i}}_{a_{i}|x_{i}}=I_{i}$ for any $i=m+1,\cdots,t$.
This implies  $\sigma_{1,\cdots, m, t+1,\cdots,n}\in\mathcal{US}_{(1\cdot\cdot\cdot m)\rightarrow (t+1\cdot\cdot\cdot n)}$.

By \cite{RAE}, for any $\rho_{12}\in\mathcal{S}(H_{1}\otimes H_{2})$ and $\rho_{1}={\rm Tr}_{2}(\rho_{12})\in\mathcal{S}(H_{1})$, we have $\|\rho_{1}\|_{1}\leq\|\rho_{12}\|_{1}$. Thus,  by Eq.(\ref{eq4.4}), there exists $\sigma_{1,\ldots, t, t+1,\ldots,n}^{(0)}\in\mathcal{US}_{(1\cdot\cdot\cdot t)\rightarrow (t+1\cdot\cdot\cdot n)}$ such that
\begin{equation}\tag{4.8}\label{eq4.7}
\begin{aligned}
&\mathbb{S}^{(1\cdot\cdot\cdot t)\rightarrow (t+1\cdot\cdot\cdot n)}(\rho_{1,\cdots, t, t+1,\cdots,n})\\=&\mathop{\min}\limits_{\sigma_{1,\cdots, t, t+1,\cdots,n}\in\mathcal{US}_{(1\cdot\cdot\cdot t)\rightarrow (t+1\cdot\cdot\cdot n)}}\frac{1}{2}\|\rho_{1,\cdots, t, t+1,\cdots,n}-\sigma_{1,\cdots, t, t+1,\cdots,n}\|_{1}\\ =&\frac{1}{2}\|\rho_{1,\cdots, t, t+1,\cdots,n}-\sigma_{1,\cdots, t, t+1,\cdots,n}^{(0)}\|_{1}\\ \geq & \frac{1}{2}\|\rho_{1,\cdots, m, t+1,\cdots,n}-\sigma_{1,\cdots, m, t+1,\cdots,n}^{(0)}\|_{1}\\ \geq &\mathop{\min}\limits_{\sigma_{1,\cdots, m, t+1,\cdots,n}\in\mathcal{US}_{(1\cdot\cdot\cdot m)\rightarrow (t+1\cdot\cdot\cdot n)}}\frac{1}{2}\|\rho_{1,\cdots, m, t+1,\cdots,n}-\sigma_{1,\cdots, m, t+1,\cdots,n}\|_{1}\\= &\mathbb{S}^{(1\cdot\cdot\cdot m)\rightarrow (t+1\cdot\cdot\cdot n)}(\rho_{1,\cdots, m, t+1,\cdots,n}),
\end{aligned}
\end{equation}
where $\sigma_{1,\cdots, m, t+1,\cdots,n}^{(0)}={\rm Tr}_{\{m+1,\cdots,t\}}(\sigma_{1,\cdots, t, t+1,\cdots,n}^{(0)})\in\mathcal{US}_{(1\cdot\cdot\cdot m)\rightarrow (t+1\cdot\cdot\cdot n)}$.

{\bf Case (2)}. $P'\preccurlyeq^{b} P$.

Without loss of generality, assume  that $P=\{1,2,\cdots,t\}$ and  $P'=P_1'|P_2'|\cdots |P_r'$ is a $r$-partition of  $P$ with $r\leq t$, where $P_{i}'=\cup_{k=1}^{s_{i}}j_{k,i}$ for $i=1,2,\cdots,r$. Additionally, we may set $Q'=Q=\{t+1,t+2,\cdots,n\}$.
 It suffices to prove the inequality
$$\mathbb{S}^{(1\cdot\cdot\cdot t)\rightarrow (t+1\cdot\cdot\cdot n)}(\rho_{1,\cdots, t, t+1,\cdots,n},N)\leq \mathbb{S}^{P'\rightarrow (t+1\cdot\cdot\cdot n)}(\rho_{1,\cdots, t, t+1,\cdots,n},M),$$
where $N=P|Q=\{1,2,\cdots,n\}$ and $M=P'|Q$.

For any $\sigma_{1,\cdots, t, t+1,\cdots,n}\in\mathcal{US}_{P'\rightarrow (t+1\cdot\cdot\cdot n)}$ and measurement assemblages $\{M^{A_{P_{1}'}}_{a_{P_{1}'}|x_{P_{1}'}}\}_{a_{P_{1}'},x_{P_{1}'}}$, $\{M^{A_{P_{2}'}}_{a_{P_{2}'}|x_{P_{2}'}}\}_{a_{P_{2}'},x_{P_{2}'}}$,$\cdot\cdot\cdot$, $\{M^{A_{P_{r}'}}_{a_{P_{r}'}|x_{P_{r}'}}\}_{a_{P_{r}'},x_{P_{r}'}}$ of $A_{P_{1}'}, A_{P_{2}'},\cdot\cdot\cdot,A_{P_{r}'}$, Eqs.(\ref{eq4.1}) and (\ref{eq4.2}) imply
\begin{equation}\tag{4.9}\label{eq4.8}
\begin{aligned}
&{\rm Tr}_{\{P_{1}',\cdots,P_{r}'\}}\{[M^{A_{P_{1}'}}_{a_{P_{1}'}|x_{P_{1}'}}\otimes\cdots\otimes M^{A_{P_{r}'}}_{a_{P_{r}'}|x_{P_{r}'}}\otimes I_{t+1}\otimes\cdots\otimes I_{n}]\sigma_{1,\cdots, t, t+1,\cdots,n}\}\\ =&\sum_{\lambda}P(\lambda)\Pi_{i=1}^{r}P(a_{P_{i}'}|x_{P_{i}'},\lambda)\sigma_{\lambda}^{t+1}\otimes\sigma_{\lambda}^{t+2}\otimes\cdot\cdot\cdot\otimes\sigma_{\lambda}^{n}.
\nonumber
\end{aligned}
\end{equation}
Then, for any  measurement assemblages $\{M^{A_{1}}_{a_{1}|x_{1}}\}_{a_{1},x_{1}}$, $\{M^{A_{2}}_{a_{2}|x_{2}}\}_{a_{2},x_{2}}$,$\cdot\cdot\cdot$, $\{M^{A_{t}}_{a_{t}|x_{t}}\}_{a_{t},x_{t}}$ of $A_{1}, A_{2},\cdot\cdot\cdot,A_{t}$,

\begin{equation}\tag{4.10}\label{eq4.9}
\begin{aligned}
&{\rm Tr}_{\{1,\cdots,t\}}\{[M^{A_{1}}_{a_{1}|x_{1}}\otimes\cdots\otimes M^{A_{t}}_{a_{t}|x_{t}}\otimes I_{t+1}\otimes\cdots\otimes I_{n}]\sigma_{1,\cdots, t, t+1,\cdots,n}\}\\ =&\sum_{\lambda}P(\lambda)\Pi_{i=1}^{r}P(a_{j_{1,i}},a_{j_{2,i}},\cdots,a_{j_{s_{i},i}}|x_{j_{1,i}},x_{j_{2,i}},\cdots,x_{j_{s_{i},i}},\lambda)\sigma_{\lambda}^{t+1}\otimes\sigma_{\lambda}^{t+2}\otimes\cdot\cdot\cdot\otimes\sigma_{\lambda}^{n}\\ =&\sum_{\lambda}P(\lambda)\Pi_{i=1}^{t}P(a_{i}|x_{i},\lambda)\sigma_{\lambda}^{t+1}\otimes\sigma_{\lambda}^{t+2}\otimes\cdot\cdot\cdot\otimes\sigma_{\lambda}^{n}.
\nonumber
\end{aligned}
\end{equation}
The first equality in inequality \eqref{eq4.9} follows from Eq.\eqref{eq4.8} by setting $M^{A_{P_{i}'}}_{a_{P_{i}'}|x_{P_{i}'}}=\otimes_{k=1}^{s_{i}} M^{A_{j_{k,i}}}_{a_{j_{k,i}}|x_{j_{k,i}}}$ for $i=1,\cdots,r$. By \eqref{eq4.2}, this implies $\sigma_{1,\cdots,t,t+1,\cdots,n} \in \mathcal{US}_{(1,\cdots,t) \to (t+1\cdots n)}$, and thus
$$\mathcal{US}_{P'\rightarrow (t+1\cdot\cdot\cdot n)}\subseteq\mathcal{US}_{(1,\cdots, t)\rightarrow (t+1\cdot\cdot\cdot n)}.$$

Consequently, by Eq.(\ref{eq4.4}), there exists $\sigma^{(0)}\in\mathcal{US}_{P'\rightarrow (t+1\cdot\cdot\cdot n)}\subseteq\mathcal{US}_{(1,\cdots, t)\rightarrow (t+1\cdot\cdot\cdot n)}$ such that
\begin{equation}\tag{4.11}\label{eq4.10}
\begin{aligned}
&\mathbb{S}^{P'\rightarrow (t+1\cdot\cdot\cdot n)}(\rho_{1,2,\cdots,n},M)\\=& \mathop{\min}\limits_{\sigma\in\mathcal{US}_{P'\rightarrow (t+1\cdot\cdot\cdot n)}}\frac{1}{2}\|\rho_{1,2,\cdots,n}-\sigma\|_{1}\\=&\frac{1}{2}\|\rho_{1,2,\cdots,n}-\sigma^{(0)}\|_{1}
\\ \geq& \mathop{\min}\limits_{\sigma\in\mathcal{US}_{(1,\cdots, t)\rightarrow (t+1\cdot\cdot\cdot n)}}\frac{1}{2}\|\rho_{1,2,\cdots,n}-\sigma\|_{1}\\=&
\mathbb{S}^{(1,\cdots, t)\rightarrow (t+1\cdot\cdot\cdot n)}(\rho_{1,2,\cdots,n}, N).
\end{aligned}
\end{equation}

{\bf Case (3)}. $P\preccurlyeq^{c} P'$.

Again, without loss of generality, we may assume  $P'=P_1'|P_2'|\cdots |P_r'$ is an $r$-partition of $\{1,2,\cdots,t\}$, $P=P_1|P_2|\cdots |P_r$ such that $P_j\subseteq P_j'$ for each $j=1,2,\cdots,r$ and $Q'=Q=Q_1|Q_2|\cdots |Q_s$ is an $s$-partition of $\{t+1,t+2,\cdots,n\}$.  Then, for any $\rho=\rho_{1,2,\cdots,n}\in{\mathcal S}(H_1\otimes H_2\otimes\cdots\otimes H_n)$, we  regard $\rho_{M}=\rho\in{\mathcal S}(H_{P_1'}\otimes H_{P_2'}\otimes\cdots\otimes H_{P_r'}\otimes H_{Q_1}\otimes H_{Q_2}\otimes\cdots\otimes H_{Q_s})$ and $\rho_N={\rm Tr}_{N^c}(\rho)\in{\mathcal S}(H_{P_1}\otimes H_{P_2}\otimes\cdots\otimes H_{P_r}\otimes H_{Q_1}\otimes H_{Q_2}\otimes\cdots\otimes H_{Q_s})$  as   $(r+s)$-partite states with respect to $M=P'|Q$ and $N=P|Q$, respectively. We need to prove the inequality
$$\mathbb{S}^{P\rightarrow Q}(\rho_N, N)\leq\mathbb{S}^{P'\rightarrow Q}(\rho_M, M).$$

For any $\sigma_{M}\in\mathcal{US}_{P'\rightarrow Q}$ and measurement assemblages $\{M^{A_{P_{1}'}}_{a_{P_{1}'}|x_{P_{1}'}}\}_{a_{P_{1}'},x_{P_{1}'}}$, $\{M^{A_{P_{2}'}}_{a_{P_{2}'}|x_{P_{2}'}}\}_{a_{P_{2}'},x_{P_{2}'}}$,$\cdot\cdot\cdot$, $\{M^{A_{P_{r}'}}_{a_{P_{r}'}|x_{P_{r}'}}\}_{a_{P_{r}'},x_{P_{r}'}}$ of $A_{P_{1}'}, A_{P_{2}'},\cdot\cdot\cdot,A_{P_{r}'}$, Eqs.(\ref{eq4.1}) and (\ref{eq4.2}) yield:
\begin{equation}\tag{4.12}\label{eq4.11}
\begin{aligned}
&{\rm Tr}_{\{P_{1}',\cdots,P_{r}'\}}\{[M^{A_{P_{1}'}}_{a_{P_{1}'}|x_{P_{1}'}}\otimes\cdots\otimes M^{A_{P_{r}'}}_{a_{P_{r}'}|x_{P_{r}'}}\otimes I_{Q_{1}}\otimes\cdots\otimes I_{Q_{s}}]\sigma_{M}\}\\ =&\sum_{\lambda}P(\lambda)\Pi_{i=1}^{r}P(a_{P_{i}'}|x_{P_{i}'},\lambda)\sigma_{\lambda}^{Q_{1}}\otimes\sigma_{\lambda}^{Q_{2}}\otimes\cdot\cdot\cdot\otimes\sigma_{\lambda}^{Q_{s}}.
\nonumber
\end{aligned}
\end{equation}

For $\sigma_{N}={\rm Tr}_{N^c}(\sigma_{M})={\rm Tr}_{\{P'\backslash P\}}(\sigma_{M})$ and any measurement assemblages $\{M^{A_{P_{1}}}_{a_{P_{1}}|x_{P_{1}}}\}_{a_{P_{1}},x_{P_{1}}}$, $\{M^{A_{P_{2}}}_{a_{P_{2}}|x_{P_{2}}}\}_{a_{P_{2}},x_{P_{2}}}$,$\cdot\cdot\cdot$, $\{M^{A_{P_{r}}}_{a_{P_{r}}|x_{P_{r}}}\}_{a_{P_{r}},x_{P_{r}}}$ of $A_{P_{1}}, A_{P_{2}},\cdot\cdot\cdot,A_{P_{r}}$,  Eq.(\ref{eq4.11}) implies:

\begin{equation}\tag{4.13}\label{eq4.12}
\begin{aligned}
&{\rm Tr}_{\{P_{1},\cdots,P_{r}\}}\{[M^{A_{P_{1}}}_{a_{P_{1}}|x_{P_{1}}}\otimes\cdots\otimes M^{A_{P_{r}}}_{a_{P_{r}}|x_{P_{r}}}\otimes I_{Q_{1}}\otimes\cdots\otimes I_{Q_{s}}]\sigma_{N}\}\\ =&{\rm Tr}_{\{P_{1},\cdots,P_{r}\}}\{[M^{A_{P_{1}}}_{a_{P_{1}}|x_{P_{1}}}\otimes\cdots\otimes M^{A_{P_{r}}}_{a_{P_{r}}|x_{P_{r}}}\otimes I_{Q_{1}}\otimes\cdots\otimes I_{Q_{s}}]{\rm Tr}_{\{P'\backslash P\}}(\sigma_{M})\}\\ =&{\rm Tr}_{\{P'\backslash P\}}\{{\rm Tr}_{\{P_{1},\cdots,P_{r}\}}[((M^{A_{P_{1}}}_{a_{P_{1}}|x_{P_{1}}}\otimes I_{\{P_{1}'\backslash P_{1}\}})\otimes\cdots\otimes (M^{A_{P_{r}}}_{a_{P_{r}}|x_{P_{r}}}\otimes I_{\{P_{r}'\backslash P_{r}\}})\otimes I_{Q_{1}}\otimes\cdots\otimes I_{Q_{s}})\sigma_{M}]\}\\ =&\sum_{\lambda}P(\lambda)\Pi_{i=1}^{r}P(a_{P_{i}}|x_{P_{i}},\lambda)\sigma_{\lambda}^{Q_{1}}\otimes\sigma_{\lambda}^{Q_{2}}\otimes\cdot\cdot\cdot\otimes\sigma_{\lambda}^{Q_{s}}.
\nonumber
\end{aligned}
\end{equation}
The final equality in inequality \eqref{eq4.12} follows from Eq.\eqref{eq4.11} by setting $M^{A_{P_{i}'}}_{a_{P_{i}'}|x_{P_{i}'}}=M^{A_{P_{i}}}_{a_{P_{i}}|x_{P_{i}}}\otimes I_{\{P_{i}'\backslash P_{i}\}}$ for $i=1,\cdots,r$.
Thus, $\sigma_{N}={\rm Tr}_{\{P'\backslash P\}}(\sigma_{M})\in\mathcal{US}_{P\rightarrow Q}$.

Consequently, by Eq.(\ref{eq4.4}), there exists $\sigma^{(0)}\in\mathcal{US}_{P'\rightarrow Q}$ such that
\begin{equation}\tag{4.14}\label{eq4.13}
\begin{aligned}
&\mathbb{S}^{P'\rightarrow Q}(\rho_{M},M)\\=&\mathop{\min}\limits_{\sigma\in\mathcal{US}_{P'\rightarrow Q}}\frac{1}{2}\|\rho_{1,2,\cdots,n}-\sigma\|_{1}\\=&\frac{1}{2}\|\rho_{M}-\sigma^{(0)}\|_{1}\\ \geq&\frac{1}{2}\|\rho_{N}-\sigma^{(0)}_{N}\|_{1}
\\ \geq&\mathop{\min}\limits_{\sigma\in\mathcal{US}_{P\rightarrow Q}}\frac{1}{2}\|\rho_{N}-\sigma\|_{1}\\=&
\mathbb{S}^{P\rightarrow Q}(\rho_{N}, N),
\end{aligned}
\end{equation}
where $\sigma^{(0)}_{N}={\rm Tr}_{N^c}(\sigma^{(0)})\in\mathcal{US}_{P\rightarrow Q}$.

Finally, combining Eqs.(\ref{eq4.7}), (\ref{eq4.10}) and (\ref{eq4.13}), if $P\preccurlyeq^{x} P'$ for $x\in\{a,c\}$ or $P'\preccurlyeq^{b} P$,  we obtain
$$\mathbb{S}^{P\rightarrow Q}(\rho_{N}, N)\leq \mathbb{S}^{P'\rightarrow Q}(\rho_{M},M),$$
completing the proof of Proposition 4.3\hfill$\Box$

{\bf Proposition 4.4.} For any $\rho_{1,2,\cdots,n}\in{\mathcal S}(H_1\otimes H_2\otimes\cdots\otimes H_n)$, and  any sub-repartitions $P'=P=P_1|P_2|\cdots|P_m\in\mathcal{SP}_{(1\rightarrow t)}$,  $Q=Q_1|Q_2|\cdots|Q_s,\ Q'=Q_1'|Q_2'|\cdots |Q_l'\in\mathcal{SP}_{(t+1\rightarrow n)}$, if $Q\preccurlyeq^{y} Q'$ for $y\in\{a,b,c\}$,  we have
$$\mathbb{S}^{P\rightarrow Q}(\rho_{N}, N)\leq \mathbb{S}^{P\rightarrow Q'}(\rho_{M},M),$$
 where $M=P'|Q'=P|Q'$, $N=P|Q$,  $\rho_{M}={\rm Tr}_{M^c}(\rho_{1,2,\cdots,n})$ and $\rho_{N}={\rm Tr}_{N^c}(\rho_{1,2,\cdots,n})$  are the reduced states of $\rho_{1,2,\cdots,n}$ to the subsystems $M$ and $N$, respectively.

{\bf Proof.} We complete the proof by considering three cases.

{\bf Case (1)}. $Q\preccurlyeq^{a}Q'$.

Without loss of generality, assume $P=P'=\{1,2,\cdots,t\}$, $Q'=\{t+1,t+2,\cdots,n\}$ and $Q=\{t+1,t+2,\cdots,t+r\}$ with $r\leq n-t$. It suffices to prove the inequality
$$\mathbb{S}^{(1\cdot\cdot\cdot t)\rightarrow (t+1\cdot\cdot\cdot t+r)}(\rho_{1,\cdots, t, t+1,\cdots,t+r})\leq \mathbb{S}^{(1\cdot\cdot\cdot t)\rightarrow (t+1\cdot\cdot\cdot n)}(\rho_{1,2,\cdots,n}),$$
where $\rho_{1,\cdots, t, t+1,\cdots,t+r}={\rm Tr}_{\{t+r+1,\cdots, n\}}(\rho_{1,2,\cdots,n})$.

For any $\sigma_{1,2,\cdots,n}\in\mathcal{US}_{(1\cdot\cdot\cdot t)\rightarrow (t+1\cdot\cdot\cdot n)}$ and  measurement assemblages $\{M^{A_{1}}_{a_{1}|x_{1}}\}_{a_{1},x_{1}}$, $\{M^{A_{2}}_{a_{2}|x_{2}}\}_{a_{2},x_{2}}$,$\cdot\cdot\cdot$, $\{M^{A_{t}}_{a_{t}|x_{t}}\}_{a_{t},x_{t}}$ of $A_{1}, A_{2},\cdot\cdot\cdot,A_{t}$, Eqs.(\ref{eq4.1}) and (\ref{eq4.2})  imply:
$$\begin{array}{rl}
&{\rm Tr}_{\{1,\cdots,t\}}\{[M^{A_{1}}_{a_{1}|x_{1}}\otimes\cdots\otimes M^{A_{t}}_{a_{t}|x_{t}}\otimes I_{t+1}\otimes\cdots\otimes I_{n}]\sigma_{1,2,\cdots,n}\}\nonumber\\ =&\sum_{\lambda}P(\lambda)\Pi_{i=1}^{t}P(a_{i}|x_{i},\lambda)\sigma_{\lambda}^{t+1}\otimes\sigma_{\lambda}^{t+2}\otimes\cdot\cdot\cdot\otimes\sigma_{\lambda}^{n}.
\nonumber
\end{array}$$

Now, consider $\sigma_{1,\cdots, t, t+1,\cdots,t+r}={\rm Tr}_{\{t+r+1,\cdots, n\}}(\sigma_{1,2,\cdots,n})$. For  any measurement assemblages $\{M^{A_{1}}_{a_{1}|x_{1}}\}_{a_{1},x_{1}}$, $\{M^{A_{2}}_{a_{2}|x_{2}}\}_{a_{2},x_{2}}$,$\cdot\cdot\cdot$, $\{M^{A_{t}}_{a_{t}|x_{t}}\}_{a_{t},x_{t}}$ of $A_{1}, A_{2},\cdot\cdot\cdot,A_{t}$,
$$\begin{array}{rl}
&{\rm Tr}_{\{1,\cdots,t\}}\{(M^{A_{1}}_{a_{1}|x_{1}}\otimes\cdots\otimes M^{A_{t}}_{a_{t}|x_{t}}\otimes I_{t+1}\otimes\cdot\cdot\cdot\otimes I_{t+r})\sigma_{1,\cdots, t, t+1,\cdots,t+r}\}\nonumber\\ =&{\rm Tr}_{\{1,\cdots,t\}}\{(M^{A_{1}}_{a_{1}|x_{1}}\otimes\cdots\otimes M^{A_{t}}_{a_{t}|x_{t}}\otimes  I_{t+1}\otimes\cdot\cdot\cdot\otimes I_{t+r}){\rm Tr}_{\{t+r+1,\cdots, n\}}(\sigma_{1,2,\cdots,n})\}\nonumber\\ =&{\rm Tr}_{\{t+r+1,\cdots, n\}}\{ {\rm Tr}_{\{1,\cdots,t\}}[(M^{A_{1}}_{a_{1}|x_{1}}\otimes\cdots\otimes M^{A_{t}}_{a_{t}|x_{t}}\otimes I_{t+1}\otimes\cdots\otimes I_{n})\sigma_{1,2,\cdots,n}]\}\nonumber\\ =&\sum_{\lambda}P(\lambda)\Pi_{i=1}^{t}P(a_{i}|x_{i},\lambda){\rm Tr}_{\{t+r+1,\cdots, n\}}\{\sigma_{\lambda}^{t+1}\otimes\sigma_{\lambda}^{t+2}\otimes\cdot\cdot\cdot\otimes\sigma_{\lambda}^{n}\}\nonumber\\ =&\sum_{\lambda}P(\lambda)\Pi_{i=1}^{t}P(a_{i}|x_{i},\lambda)\sigma_{\lambda}^{t+1}\otimes\sigma_{\lambda}^{t+2}\otimes\cdot\cdot\cdot\otimes\sigma_{\lambda}^{t+r}.
\nonumber
\end{array}$$
This establishes that $\sigma_{1,\cdots, r, t+1,\cdots,t+r}\in\mathcal{US}_{(1\cdot\cdot\cdot t)\rightarrow (t+1\cdot\cdot\cdot t+r)}$.

Note that $\|\rho_{1}\|_{1}\leq\|\rho_{12}\|_{1}$ holds  for any $\rho_{12}\in\mathcal{S}(H_{1}\otimes H_{2})$ and $\rho_{1}={\rm Tr}_{2}(\rho_{12})\in\mathcal{S}(H_{1})$ by  \cite{RAE}.  This directly implies the existence of a state  $\sigma_{1,2,\cdots,n}^{(0)}\in\mathcal{US}_{(1\cdot\cdot\cdot t)\rightarrow (t+1\cdot\cdot\cdot n)}$ such that
\begin{equation}\tag{4.15}\label{eq4.14}
\begin{aligned}
&\mathbb{S}^{(1\cdot\cdot\cdot t)\rightarrow (t+1\cdot\cdot\cdot n)}(\rho_{1,2,\cdots,n})\\ =&\mathop{\min}\limits_{\sigma_{1,2,\cdots,n}\in\mathcal{US}_{(1\cdot\cdot\cdot t)\rightarrow (t+1\cdot\cdot\cdot n)}}\frac{1}{2}\|\rho_{1,2,\cdots,n}-\sigma_{1,2,\cdots,n}\|_{1}\\ =&\frac{1}{2}\|\rho_{1,2,\cdots,n}-\sigma_{1,2,\cdots,n}^{(0)}\|_{1}
\nonumber\\ \geq&\frac{1}{2}\|\rho_{1,\cdots, t, t+1,\cdots,t+r}-\sigma_{1,\cdots, t, t+1,\cdots,t+r}^{(0)}\|_{1}\\ \geq&\mathop{\min}\limits_{\sigma_{1,\cdots, t, t+1,\cdots,t+r}\in\mathcal{US}_{(1\cdot\cdot\cdot t)\rightarrow (t+1\cdot\cdot\cdot t+r)}}\frac{1}{2}\|\rho_{1,\cdots, t, t+1,\cdots,t+r}-\sigma_{1,\cdots, t, t+1,\cdots,t+r}\|_{1}\\ =&\mathbb{S}^{(1\cdot\cdot\cdot t)\rightarrow (t+1\cdot\cdot\cdot t+r)}(\rho_{1,\cdots, t, t+1,\cdots,t+r}),
\end{aligned}
\end{equation}
where $\rho_{1,\cdots, t, t+1,\cdots,t+r}={\rm Tr}_{\{t+r+1,\cdots, n\}}(\rho_{1,2,\cdots,n})$ and $\sigma_{1,\cdots, t, t+1,\cdots,t+r}^{(0)}={\rm Tr}_{\{t+r+1,\cdots, n\}}(\sigma_{1,2,\cdots,n}^{(0)})\in\mathcal{US}_{(1\cdot\cdot\cdot t)\rightarrow (t+1\cdot\cdot\cdot t+r)}$.

{\bf Case (2)}. $Q\preccurlyeq^{b}Q'$.

Assume without loss of generality that $P=P'=\{1,2,\cdots,t\}$, $Q'=\{t+1,t+2,\cdots,n\}$, and  $Q$ is an $s$-partition of  $Q'$ with $s<n-t$. It suffices to prove the inequality
$$\mathbb{S}^{(1\cdot\cdot\cdot t)\rightarrow Q}(\rho_{1,2,\cdots,n}, N)\leq \mathbb{S}^{(1\cdot\cdot\cdot t)\rightarrow (t+1\cdot\cdot\cdot n)}(\rho_{1,2,\cdots,n},M),$$
where $M=P'|Q'=\{1,2,\cdots,n\}$, $N=P|Q$.

For any $\sigma_{1,2,\cdots,n}\in\mathcal{US}_{(1\cdot\cdot\cdot t)\rightarrow (t+1\cdot\cdot\cdot n)}$ and measurement assemblages $\{M^{A_{1}}_{a_{1}|x_{1}}\}_{a_{1},x_{1}}$, $\{M^{A_{2}}_{a_{2}|x_{2}}\}_{a_{2},x_{2}}$,$\cdot\cdot\cdot$,  $\{M^{A_{t}}_{a_{t}|x_{t}}\}_{a_{t},x_{t}}$ of $A_{1}, A_{2},\cdot\cdot\cdot,A_{t}$, Eqs.(\ref{eq4.1}) and (\ref{eq4.2}) imply:
$$\begin{array}{rl}
\sigma_{a_{1},\cdots,a_{t}|x_{1},\cdots,x_{t}}=&{\rm Tr}_{\{1,\cdots,t\}}\{[M^{A_{1}}_{a_{1}|x_{1}}\otimes\cdots\otimes M^{A_{t}}_{a_{t}|x_{t}}\otimes I_{t+1}\otimes\cdots\otimes I_{n}]\sigma_{1,2,\cdots,n}\}\nonumber\\ =&\sum_{\lambda}P(\lambda)\Pi_{i=1}^{t}P(a_{i}|x_{i},\lambda)\sigma_{\lambda}^{t+1}\otimes\sigma_{\lambda}^{t+2}\otimes\cdot\cdot\cdot\otimes\sigma_{\lambda}^{n}.
\nonumber
\end{array}$$
Define  $\sigma_{\lambda}^{Q_{j}}=\otimes_{i\in Q_{j}}\sigma_{\lambda}^{i}$ for any $j=1,2,\cdots,s$, then
$$\begin{array}{rl}
\sigma_{a_{1},\cdots,a_{t}|x_{1},\cdots,x_{t}} =&\sum_{\lambda}P(\lambda)\Pi_{i=1}^{t}P(a_{i}|x_{i},\lambda)\sigma_{\lambda}^{t+1}\otimes\sigma_{\lambda}^{t+2}\otimes\cdot\cdot\cdot\otimes\sigma_{\lambda}^{n}.
\nonumber\\=&\sum_{\lambda}P(\lambda)\Pi_{i=1}^{t}P(a_{i}|x_{i},\lambda)\sigma_{\lambda}^{Q_{1}}\otimes\cdot\cdot\cdot\otimes\sigma_{\lambda}^{Q_{s}}\nonumber.
\end{array}$$
By Eq.(\ref{eq4.2}), this implies $\sigma_{1,2,\cdots,n}\in\mathcal{US}_{(1\cdot\cdot\cdot t)\rightarrow Q}$, and consequently
$$\mathcal{US}_{(1\cdot\cdot\cdot t)\rightarrow (t+1\cdot\cdot\cdot n)}\subseteq \mathcal{US}_{(1\cdot\cdot\cdot t)\rightarrow Q}.$$
Therefore, by Eq.(\ref{eq4.4}), there exists $\sigma^{(0)}\in\mathcal{US}_{(1\cdot\cdot\cdot t)\rightarrow (t+1\cdot\cdot\cdot n)}\subseteq \mathcal{US}_{(1\cdot\cdot\cdot t)\rightarrow Q}$ such that
\begin{equation}\tag{4.16}\label{eq4.15}
\begin{aligned}
&\mathbb{S}^{(1\cdot\cdot\cdot t)\rightarrow (t+1\cdot\cdot\cdot n)}(\rho_{1,2,\cdots,n},M)\\=&\mathop{\min}\limits_{\sigma\in\mathcal{US}_{(1\cdot\cdot\cdot t)\rightarrow (t+1\cdot\cdot\cdot n)}}\frac{1}{2}\|\rho_{1,2,\cdots,n}-\sigma\|_{1}\\=&\frac{1}{2}\|\rho_{1,2,\cdots,n}-\sigma^{(0)}\|_{1}
\\ \geq&\mathop{\min}\limits_{\sigma\in\mathcal{US}_{(1\cdot\cdot\cdot t)\rightarrow Q}}\frac{1}{2}\|\rho_{1,2,\cdots,n}-\sigma\|_{1}\\=&
\mathbb{S}^{(1\cdot\cdot\cdot t)\rightarrow Q}(\rho_{1,2,\cdots,n}, N).
\end{aligned}
\end{equation}

{\bf Cases(3)}. $Q\preccurlyeq^{c}Q'$.

Also, we may assume without loss of generality that $Q'=Q_1'|Q_2'|\cdots |Q_l'$ is an $l$-partition of $\{t+1,t+2,\cdots,n\}$ and $Q=Q_1|Q_2|\cdots |Q_l$ satisfies $Q_i\subseteq Q_i'$ for  $i=1,2,\cdots,l$. Furthermore, assume $P'=P=P_1|P_2|\cdots |P_m$ be an $m$-partition of $\{1,2,\cdots,t\}$.  Then, for any $\rho=\rho_{1,2,\cdots,n}\in{\mathcal S}(H_1\otimes H_2\otimes\cdots\otimes H_n)$, we  regard $\rho_{M}=\rho\in{\mathcal S}(H_{P_1}\otimes H_{P_2}\otimes\cdots\otimes H_{P_m}\otimes H_{Q_1'}\otimes H_{Q_2'}\otimes\cdots\otimes H_{Q_l'})$ and $\rho_N={\rm Tr}_{N^c}(\rho)\in{\mathcal S}(H_{P_1}\otimes H_{P_2}\otimes\cdots\otimes H_{P_m}\otimes H_{Q_1}\otimes H_{Q_2}\otimes\cdots\otimes H_{Q_l})$  as   $(m+l)$-partite states with respect to $M=P|Q'$ and $N=P|Q$, respectively. We need to prove the inequality
$$\mathbb{S}^{P\rightarrow Q}(\rho_N, N)\leq\mathbb{S}^{P\rightarrow Q'}(\rho_M, M).$$

For any $\sigma_{M}\in\mathcal{US}_{P\rightarrow Q'}$,  Eqs.(\ref{eq4.1}) and (\ref{eq4.2}) imply that for  any measurement assemblages $\{M^{A_{P_{1}}}_{a_{P_{1}}|x_{P_{1}}}\}_{a_{P_{1}},x_{P_{1}}}$, $\{M^{A_{P_{2}}}_{a_{P_{2}}|x_{P_{2}}}\}_{a_{P_{2}},x_{P_{2}}}$,$\cdot\cdot\cdot$, $\{M^{A_{P_{m}}}_{a_{P_{m}}|x_{P_{m}}}\}_{a_{P_{m}},x_{P_{m}}}$ of $A_{P_{1}}, A_{P_{2}},\cdot\cdot\cdot,A_{P_{m}}$,
\begin{equation}\tag{4.17}\label{eq4.16}
\begin{aligned}
&{\rm Tr}_{\{P_{1},\cdots,P_{m}\}}\{[M^{A_{P_{1}}}_{a_{P_{1}}|x_{P_{1}}}\otimes\cdots\otimes M^{A_{P_{m}}}_{a_{P_{m}}|x_{P_{m}}}\otimes I_{Q_{1}'}\otimes\cdots\otimes I_{Q_{l}'}]\sigma_{M}\}\\ =&\sum_{\lambda}P(\lambda)\Pi_{i=1}^{m}P(a_{P_{i}}|x_{P_{i}},\lambda)\sigma_{\lambda}^{Q_{1}'}\otimes\sigma_{\lambda}^{Q_{2}'}\otimes\cdot\cdot\cdot\otimes\sigma_{\lambda}^{Q_{l}'}.
\nonumber
\end{aligned}
\end{equation}

Now consider $\sigma_{N}={\rm Tr}_{N^c}(\sigma_{M})={\rm Tr}_{\{Q'\backslash Q\}}(\sigma_{M})$. For any measurement assemblages $\{M^{A_{P_{1}}}_{a_{P_{1}}|x_{P_{1}}}\}_{a_{P_{1}},x_{P_{1}}}$, $\{M^{A_{P_{2}}}_{a_{P_{2}}|x_{P_{2}}}\}_{a_{P_{2}},x_{P_{2}}}$,$\cdot\cdot\cdot$, $\{M^{A_{P_{m}}}_{a_{P_{m}}|x_{P_{m}}}\}_{a_{P_{m}},x_{P_{m}}}$ of $A_{P_{1}}, A_{P_{2}},\cdot\cdot\cdot,A_{P_{m}}$,  Eq.(\ref{eq4.16}) directly implies that

\begin{equation}\tag{4.18}\label{eq4.17}
\begin{aligned}
&{\rm Tr}_{\{P_{1},\cdots,P_{m}\}}\{[M^{A_{P_{1}}}_{a_{P_{1}}|x_{P_{1}}}\otimes\cdots\otimes M^{A_{P_{m}}}_{a_{P_{m}}|x_{P_{m}}}\otimes I_{Q_{1}}\otimes\cdots\otimes I_{Q_{l}}]\sigma_{N}\}\\ =&{\rm Tr}_{\{P_{1},\cdots,P_{m}\}}\{[M^{A_{P_{1}}}_{a_{P_{1}}|x_{P_{1}}}\otimes\cdots\otimes M^{A_{P_{m}}}_{a_{P_{m}}|x_{P_{m}}}\otimes I_{Q_{1}}\otimes\cdots\otimes I_{Q_{l}}]{\rm Tr}_{\{Q'\backslash Q\}}(\sigma_{M})\}\\ =&{\rm Tr}_{\{Q'\backslash Q\}}\{{\rm Tr}_{\{P_{1},\cdots,P_{m}\}}[(M^{A_{P_{1}}}_{a_{P_{1}}|x_{P_{1}}}\otimes\cdots\otimes M^{A_{P_{m}}}_{a_{P_{m}}|x_{P_{m}}}\otimes I_{Q_{1}'}\otimes\cdots\otimes I_{Q_{l}'})\sigma_{M}]\}\\ =&\sum_{\lambda}P(\lambda)\Pi_{i=1}^{m}P(a_{P_{i}}|x_{P_{i}},\lambda){\rm Tr}_{\{Q'\backslash Q\}}\{\sigma_{\lambda}^{Q_{1}'}\otimes\sigma_{\lambda}^{Q_{2}'}\otimes\cdot\cdot\cdot\otimes\sigma_{\lambda}^{Q_{l}'}\}\\ =&\sum_{\lambda}P(\lambda)\Pi_{i=1}^{m}P(a_{P_{i}}|x_{P_{i}},\lambda)\sigma_{\lambda}^{Q_{1}}\otimes\sigma_{\lambda}^{Q_{2}}\otimes\cdot\cdot\cdot\otimes\sigma_{\lambda}^{Q_{l}}.
\nonumber
\end{aligned}
\end{equation}
Thus, $\sigma_{N}={\rm Tr}_{N^c}(\sigma_{M})\in\mathcal{US}_{P\rightarrow Q}$.

Therefore, by Eq.(\ref{eq4.4}), there exists $\sigma^{(0)}\in\mathcal{US}_{P\rightarrow Q'}$ such that
\begin{equation}\tag{4.19}\label{eq4.18}
\begin{aligned}
&\mathbb{S}^{P\rightarrow Q'}(\rho_{M},M)\\=&\mathop{\min}\limits_{\sigma\in\mathcal{US}_{P\rightarrow Q'}}\frac{1}{2}\|\rho_{1,2,\cdots,n}-\sigma\|_{1}\\=&\frac{1}{2}\|\rho_{M}-\sigma^{(0)}\|_{1}\\ \geq&\frac{1}{2}\|\rho_{N}-\sigma^{(0)}_{N}\|_{1}
\\ \geq&\mathop{\min}\limits_{\sigma\in\mathcal{US}_{P\rightarrow Q}}\frac{1}{2}\|\rho_{N}-\sigma\|_{1}\\=&
\mathbb{S}^{P\rightarrow Q}(\rho_{N}, N),
\end{aligned}
\end{equation}
where $\sigma^{(0)}_{N}={\rm Tr}_{N^c}(\sigma^{(0)})\in\mathcal{US}_{P\rightarrow Q}$.

Then, by Eqs.(\ref{eq4.14}), (\ref{eq4.15}) and (\ref{eq4.18}), if $Q\preccurlyeq^{y} Q'$ for $y\in\{a,b,c\}$,  we have
$$\mathbb{S}^{P\rightarrow Q}(\rho_{N}, N)\leq \mathbb{S}^{P\rightarrow Q'}(\rho_{M},M).$$
\hfill$\Box$

Now, by the definition of $(P,Q)\preccurlyeq_s(P',Q')$ in Definition 4.2 and Eq.(\ref{eq4.0}),  it is easily seen from  Propositions 4.3 and 4.4 that the function $\mathbb{S}^{(1\cdot\cdot\cdot t)\rightarrow (t+1\cdot\cdot\cdot n)}$ satisfies condition (MStM4).

Therefore,  $\mathbb{S}^{(1\cdot\cdot\cdot t)\rightarrow (t+1\cdot\cdot\cdot n)}$ satisfies the conditions (MStM1)-(MStM4), making it a true measure of multipartite  steering. Consequently, multipartite  steering is an asymmetric multipartite quantum resource, that is, Theorem 4.1 is true.

\section{Monogamy  relations for MQC measures}

At present, we already know that many MQCs are multipartite quantum  resources in the sense mentioned in section 2, and then they should obey further the rules of resource allocation theory such as monogamy relations. Thus, naturally, when discussing MQC measures, the monogamy relations should be explored. However, the monogamy relations of a given MQC are closely related to the hierarchy condition with respect to MQC and thus are difficult to give a uniform definition.
In this section, we provide a precise definition of the monogamy for true measures of symmetric MQCs with $\preccurlyeq$ defined in Eq.(\ref{eq3.1}) as the hierarchy relation between sub-repartitions because many symmetric MQCs take the hierarchy condition concerning $\preccurlyeq$. Furthermore, based on the three basic types (a), (b) and (c) of hierarchy  relations of sub-repartitions, we define four specific types of monogamy relations: global monogamy relation, complete monogamy relation, tight monogamy relation, and strong monogamy relation.

For convenience,  if $Q=Q_1|Q_2|\cdots |Q_r\in\mathcal{SP}_{n}$, $P=P_1|P_2|\cdots |P_m\in\mathcal{SP}_{n}$ and $Q\preccurlyeq P$, we denote by $\Xi(P-Q)$ the set of all sub-repartitions that are coarser than $P$ but (1) neither coarser than $Q$ nor the one from which we can derive $Q$ by the coarsening means; (2) if it includes some or all subsystems of $Q$, then all  subsystems $Q_{j}$s included are regarded as one subsystem, and (3) if $Q_1|Q_2|\cdots |Q_r=P_1|P_2|\cdots |P_{r-1}|P_{r}\cdots P_{m}$, $\Xi(P-Q)$ contains only $P_{r}|\cdots |P_{m}$ and  those coarser than it. We call
$\Xi(P-Q)$  {\it the complementarity of $Q$ up to $P$} \cite{GY24}. For example,
$$\begin{array}{rl}\Xi(1|2|34|5-1|2) =&\{34|5, 1|34|5, 2|34|5, 1|34, 2|34, 2|3|5, 2|4|5, 1|3|5, 1|4|5, 1|5,  2|5, \nonumber\\& 1|3, 1|4, 2|3, 2|4, 3|5, 4|5, 1|345, 2|345, 12|345, 12|34|5, 12|34, 12|5\}.\end{array}$$

Below, we provide a precise definition of the monogamy for true measures of symmetric MQCs.

{\bf Definition 5.1.} (Global monogamy relation) For $n\geq2$, assume that $\mathcal{C}^{(n)}$ is a true MQC measure for a symmetric MQC $\mathcal C$  in an $n$-partite composite system $H_1\otimes H_2\otimes\cdots\otimes H_n$ of which the  hierarchy relation between sub-repartitions associated with $\mathcal C$  is $\preccurlyeq$, that is, $\preccurlyeq_{\mathcal C}=\preccurlyeq$.  We say that $\mathcal{C}^{(n)}$ is  globally monogamous  if, for any  $P=P_1|P_2|\cdots |P_m$, $Q=Q_1|Q_2|\cdots |Q_r\in\mathcal{SP}_{n}$, and any $n$-partite state $\rho_{1,2,\cdot\cdot\cdot,n}\in\mathcal{S}(H_1\otimes H_2\otimes\cdots\otimes H_n)$,
$$Q\preccurlyeq P \quad{\rm and}\quad \mathcal{C}^{(m)}(\rho_{P})=\mathcal{C}^{(r)}(\rho_{Q})$$
will imply that
$$\mathcal{C}^{(\ast)}(\rho_{\Upsilon})=0$$
holds for any $\Upsilon\in\Xi(P-Q)$, and hereafter the superscript asterisk (*)  is associated with the partite of sub-repartition $\Upsilon$. For example, if $\Upsilon$ is a $t$-partite sub-repartition, then $\ast=t$.

Generally, the global monogamy of true MQC measures is difficult to be checked. Thus, we propose further three basic types of monogamy relations below.

{\bf Definition 5.2.} Let $\mathcal{C}^{(n)}$ be a true MQC measure for symmetric MQC $\mathcal C$ in an $n$-partite composite system $H_1\otimes H_2\otimes\cdots\otimes H_n$ ($n\geq 2$) with the hierarchy relation  $\preccurlyeq_{\mathcal C}=\preccurlyeq$.

(1) (Complete monogamy relation) $\mathcal{C}^{(n)}$ $(n\geq3)$ is completely monogamous if, for any $n$-partite state $\rho_{1,2,\cdot\cdot\cdot,n}\in\mathcal{S}(H_1\otimes H_2\otimes\cdots\otimes H_n)$ and any  $P=P_1|P_2|\cdots |P_m$, $Q=Q_1|Q_2|\cdots |Q_r\in\mathcal{SP}_{n}$,
$$Q\preccurlyeq^a P\quad{\rm and}\quad \mathcal{C}^{(m)}(\rho_{P})=\mathcal{C}^{(r)}(\rho_{Q})$$
will imply that
$$\mathcal{C}^{(2)}(\rho_{Z|W})=0$$
and if $m-r\geq2$,
$$\mathcal{C}^{(m-r)}(\rho_{\overline{W}})=0,$$
where $Z=\bigcup_{j=1}^r Q_j$, $W=\bigcup_{i=1}^m P_i\setminus Z$ and $\overline{W}=P\setminus Q$. \if false Here $W$ denotes the subsystem(s) of $P_1P_2\cdots P_m$ complementary to $Q_1Q_2\cdots Q_r$ and $\overline{W}$ denotes the complementarity of $P$ up to $Q$.\fi

For example, if $P=1|2|3|4|5$ and $Q=2|4$, then $Z=\{2,4\}$, $W=\{1,3,5\}$ and $\overline{W}=1|3|5$. Thus $\rho_{Z|W}=\rho_{24|135}$ and $\rho_{\overline{W}}=\rho_{1,3,5}$.

(2) (Tight monogamy relation) $\mathcal{C}^{(n)}$ $(n\geq3)$ is tightly monogamous if, for any $n$-partite state $\rho_{1,2,\cdot\cdot\cdot,n}\in\mathcal{S}(H_1\otimes H_2\otimes\cdots\otimes H_n)$,
$$Q\preccurlyeq^b P\quad{\rm and}\quad \mathcal{C}^{(m)}(\rho_{P})=\mathcal{C}^{(r)}(\rho_{Q})$$
will imply
$$C^{(s_{i})}(\rho_{P_{j_{1},i}|P_{j_{2},i}|\cdots|P_{j_{s_{i}},i}})=0,$$
where $ P_{j_{1},i}|P_{j_{2},i}|\cdots|P_{j_{s_{i}},i}=Q_i$.

(3) (Strong monogamy relation) $\mathcal{C}^{(n)}$ $(n\geq2)$ is strongly monogamous if, for any $n$-partite state $\rho_{1,2,\cdot\cdot\cdot,n}\in\mathcal{S}(H_1\otimes H_2\otimes\cdots\otimes H_n)$,
$$Q\preccurlyeq^c P\quad{\rm and}\quad \mathcal{C}^{(m)}(\rho_{P})=\mathcal{C}^{(m)}(\rho_{Q})$$
will imply that
$\mathcal{C}^{(i_{s})}(\rho_{Q_{i}})=0$ whenever $i_{s}\geq2$, with $i_{s}$ the number of subsystems contained in $Q_{i}$, and that $\mathcal{C}^{(2)}(\rho_{ij})=0$ whenever one of $i$ and $j$ is not in $\bigcup_{i=1}^m Q_{i}$.

Roughly speaking, {\it the complete monogamy relation} indicates that if the correlation within a subgroup of subsystems attains the total correlation of the entire system, then parties outside of this subgroup are not correlated to any other parties within the system. The concept of {\it tight monogamy relation} implies that if the correlation within a partition attains the total correlation, then the parties within the same subgroup are not correlated to each other. {\it The strong monogamy relation} asserts that if the correlation within a partition remains unchanged after removing some parties from  each subgroup, then the remaining parties are not correlated with the parties removed.

It is noteworthy that if a true MQC measure $\mathcal{C}^{(n)}$ satisfies the global monogamy relation, then $\mathcal{C}^{(n)}$ must inherently be the complete monogamy relation, tight monogamy relation, and strong monogamy relation.

We do not know if there exist true symmetric MQCs measures that are globally monogamous. But do there exist some true symmetric MQCs measures that are completely monogamous and tightly monogamous.

{\bf Example 5.1.} By \cite{GY20}, it is known that the tripartite entanglement measures $E_{f}^{(3)}, C^{(3)}$, and $T_{q}^{(3)}$ in Example 3.1 are completely monogamous and tightly monogamous.

{\bf Example 5.2.}  For the  multipartite multi-mode Gaussian non-product correlation measure $\mathcal{M}^{(n)}$ in Example 3.7,   it is known from \cite{HLQ22} that $\mathcal{M}^{(n)}$  is completely monogamous and tightly monogamous. However, $\mathcal{M}^{(n)}$ is not strongly monogamous.

{\bf Example 5.3.}  Regarding the multipartite single-mode Gaussian
coherence measure $C_\nu^{G_n}$ in Example 3.9,  the only appropriate monogamy relation  is the complete monogamy. By \cite{HHQ24}, it is known that $C_\nu^{G_n}$ is completely monogamous.

The monogamy relations for an asymmetric MQC measure can be defined similarly according to what the hierarchy relation is.


\section{Conclusion}

Currently, the fundamental elements required by multipartite quantum resource theory are consistent with those of bipartite quantum resource theory, namely free states, free operations, and resource measures. Here, resource measures should satisfy the faithfulness  and non-increasing trend under free operations.
However,  the multipartite quantum correlation (MQC) measures should satisfy more principles from many-boy resource allocation theory, making the  current quantum resource theory not applicable to MQCs. In this paper, we establish a more reasonable framework for multipartite quantum resource theory, primarily by more accurately defining the conditions that true MQC measures should satisfy. Motivated by multipartite entanglement measures \cite{GY20} and multipartite multi-mode Gaussian non-product measure \cite{HLQ22}, a true MQC measure, in addition to the usual fundamental principles such as the faithfulness (MQCM1) and monotonicity under free operations (MQCM2), should also satisfy the unification condition (MQCM3), which enables us to measure the same source contained in part systems, and the hierarchy  condition (MQCM4), which ensures that the resource hold by part system does not exceed that in entire system. We call such MQC measures the true MQC measures distinguishing from those that only satisfy (MQCM1) and (MQCM2).

\if false Specifically, we first investigate the measures of completely symmetric MQCs. In this case, for condition (MQCM4),  we provide three different levels of hierarchy  relationships (CS-MQCM4a)-(CS-MQCM4c) based on three distinct types of coarsening relations of multipartite partition. Furthermore, in addition to conditions (MQCM1)-(MQCM4), a true MQC measure for completely symmetric MQCs should also satisfy symmetry (MQCM5). Based on these principles, we find that multipartite entanglement, multipartite non-PPT, multipartite coherence, multipartite  imaginarity, multipartite multimode Gaussian non-product correlation, and multipartite multimode Gaussian imaginarity are completely symmetric multipartite quantum resources. In this scenario, two special types of completely symmetric MQCs exist: $k$-entanglement and $k$-partite entanglement for $n$-partite system. For $k$-entanglement and $k$-partite entanglement, we propose the more general unification condition  and the hierarchy condition. Based on these conditions, we illustrate that $k$-entanglement and $k$-partite entanglement are  completely symmetric  multipartite quantum resources. In addition to completely symmetric MQCs, there also exist symmetric but not completely symmetric MQCs. In such cases, a true MQC measure should satisfy conditions (MQCM1)-(MQCM3) and (MQCM5), and the hierarchy condition may be part of (CS-MQCM4a)-(CS-MQCM4c). Based on this, we demonstrate that multipartite single-mode Gaussian coherence is a symmetric multipartite Gaussian quantum resource. Furthermore, there exist asymmetric MQCs. In such cases, a true MQC measure should satisfy conditions (MQCM1)-(MQCM3), as well as the hierarchy condition determined by the multipartite quantum correlation itself. Here, we elaborate on  multipartite discord as an asymmetric multipartite quantum resource.\fi

Several  symmetric MQCs are shown to be multipartite quantum resources as at least one true MQC measure is presented, such as the entanglement, $k$-entanglement, $k$-partite entanglement, non-MPPT, multipartite coherence, multipartite imaginarity, multipartite multi-mode Gaussian non-product correlation, multipartite multi-mode Gaussian imaginarity, multipartite single-mode Gaussian coherence. Among them, the results concerning non-MPPT, multipartite coherence and multipartite imaginarity are new. For asymmetric MQC, we prove that multipartite steering  is an  asymmetric multipartite quantum resource. Note that, due to the  inherent characteristics of MQCs, the true measures of different MQCs exhibit distinct hierarchy conditions.

Finally, we discuss the monogamy relation for true measures of symmetric MQCs. Since the monogamy of true MQC measures is closely related to the hierarchy condition of the given MQC and is difficult to describe, we  revisit the monogamy relations of true MQC measures for some symmetric MQCs  with $\preccurlyeq$ as the hierarchy relation between sub-repartitions. Based on three types of coarsening relations $\preccurlyeq^x$, $x\in\{a,b,c\}$,  complete monogamy relation, tight monogamy relation, and strong monogamy relation as well as global monogamy relation are defined. Some  known true MQC measures are completely monogamous and tightly monogamous.
 It remains unclear if there are true MQC measures that are strongly monogamous or globally monogamous.

It is also interesting to describe the monogamy relation for general MQCs with hierarchy order $\preccurlyeq_{\mathcal C}$ different from $\preccurlyeq$. Particularly,  what are the monogamy relations  for true measures of $k$-entanglement, $k$-partite entanglement and multipartite steering? These topics are worth to be explored further.

{\bf Acknowledgement.}
This work was supported by the National Natural Science Foundation
 of China (Grant Nos. 12571138, 12071336, 12171290, 12271394)

\end{document}